\documentclass[twocolumn,showpacs,nofootinbib]{revtex4}
\usepackage{epsf}
\usepackage{graphicx}
\usepackage{dcolumn}
\usepackage{multirow}

\def\lsim{~\rlap{$<$}{\lower 1.0ex\hbox{$\sim$}}}
\def\gsim{~\rlap{$>$}{\lower 1.0ex\hbox{$\sim$}}}

\newcommand{\BE}{\begin{equation}}
\newcommand{\EE}{\end{equation}}
\newcommand{\BA}{\begin{eqnarray}}
\newcommand{\EA}{\end{eqnarray}}

\def\be{\begin{equation}}
\def\ee{\end{equation}}
\def\bea{\begin{eqnarray}}
\def\eea{\end{eqnarray}}

\def\fun#1#2{\lower3.6pt\vbox{\baselineskip0pt\lineskip.9pt
        \ialign{$\mathsurround=0pt#1\hfill##\hfil$\crcr#2\crcr\sim\crcr}}}

\def\apj{ApJ\,}                 
\def\apjl{ApJL\,}                
\def\apjs{ApJS\,}               
\def\caa{ChA\&A\,}
\def\mnras{MNRAS\,}             
\def\aap{A\&A\,}                
\def\physrep{Phys.~Rep.\,}   

\begin{document}
\input epsf
\renewcommand{\topfraction}{0.8}
\preprint{astro-ph/yymmnnn, \today}


\vspace{1cm}

\title{\bf\LARGE Baryon impact on weak lensing peaks and power spectrum: 
low-bias statistics and self-calibration in future surveys}

\author{\bf Xiuyuan Yang$^{1,4,5}$, Jan M. Kratochvil$^{2}$, Kevin Huffenberger$^{2}$, Zolt\'an Haiman$^{3,5}$, Morgan May$^{4}$}

\affiliation{ {$^1$ Department of Physics, Columbia University, New York, NY 10027, USA}} 
\affiliation{ {$^2$ Department of Physics, University of Miami, Coral Gables, FL 33124, USA}}
\affiliation{ {$^3$ Department of Astronomy and Astrophysics, Columbia University, New York, NY 10027, USA}} 
\affiliation{ {$^4$ Physics Department, Brookhaven National Laboratory, Upton, NY 11973, USA}   }
\affiliation{ {$^5$ Institute for Strings, Cosmology, and Astroparticle Physics (ISCAP), Columbia University, New York, NY 10027, USA}}


{\begin{abstract} Peaks in two-dimensional weak lensing (WL) maps
contain significant cosmological information, complementary to the WL
power spectrum.  This has recently been demonstrated using N-body
simulations which neglect baryonic effects. Here we employ ray-tracing
N-body simulations in which we manually steepen the density profile of
each dark matter halo, mimicking the cooling and concentration of
baryons into dark matter potential wells.  We find, in agreement with
previous works, that this causes a significant increase in the
amplitude of the WL power spectrum on small scales (spherical harmonic
index $\ell\gsim 1,000$). We then study the impact of the halo
concentration increase on the peak counts, and find the following.
(i) Low peaks (with convergence $0.02 \lsim \kappa_{\rm peak} \lsim
0.08$), remain nearly unaffected.  These peaks are created by a
constellation of several halos with low masses ($\sim 10^{12}-10^{13}
{\rm M_\odot}$) and large angular offsets from the peak center ($\gsim
0.5 R_{\rm vir}$); as a result, they are insensitive to the central
halo density profiles.  These peaks contain most of the cosmological
information, and thus provide an unusually sensitive and unbiased
probe.
(ii) The number of high peaks (with convergence $\kappa_{\rm peak}
\gsim 0.08$) is increased.  However, when the baryon effects are
neglected in cosmological parameter estimation, then the high peaks
lead to a modest bias, comparable to that from the power spectrum on
relatively large-scales ($\ell<2000$), and much smaller than the bias
from the power spectrum on smaller scales ($\ell>2,000$).
(iii) In the 3D parameter space ($\sigma_8, \Omega_m, w$), the biases
from the high peaks and the power spectra are in different directions.
This suggests the possibility of ``self-calibration'': the combination
of peak counts and power spectrum can simultaneously constrain
baryonic physics and cosmological parameters.
\end{abstract}}
\maketitle

\section{Introduction}\label{Introduction}

Weak gravitational lensing (WL) by large-scale cosmic structures has
emerged as one of the most promising methods to constrain the
parameters of both dark energy (DE) and dark matter (DM)
(e.g. ref.~\cite{DETF}; see also recent reviews in
refs.~\cite{HJ08,Munshi+08}).  WL was first detected over two decades
ago~\cite{Tysonetal}. It has recently matured to deliver cosmological
constraints; in particular, the COSMOS survey has provided independent
evidence of the accelerated expansion of the Universe~\cite{COSMOS,
Semboloni+11b}. Over the next decade, revolutionary large WL datasets
are expected to be available.  The LSST survey will cover 20,000
square degrees with multi-band imaging suitable for weak lensing, and
other large surveys will cover several thousand square
degrees~\cite{surveys}.  These WL datasets will be a rich source of
cosmological information, going beyond traditional two-point
statistics such as the power spectrum.

Lensing peaks, defined as local maxima in two-dimensional WL maps,
provide statistics that will be automatically available and
particularly straightforward to measure in lensing datasets.  Since
they probe the underlying 3D density fluctuations on small, non-linear
scales, they are sensitive to non-Gaussian features.  As a
cosmological probe, the peak counts are therefore complementary to the
WL power spectrum, and are similar to galaxy cluster counts.  A major
advantage of peaks and a motivation for their use is that they avoid
the issue of having to identify genuine bound clusters and measure
their masses.  Accurate modeling of peak statistics requires large
numerical simulations, which are now becoming feasible.

In the past few years, interest in lensing peaks and other, closely
related statistics has increased significantly.\footnote{To our
knowledge, lensing peaks were first considered as a cosmology probe in
the early ray-tracing simulations by ref.~\cite{JV00}, which studied
the $\Omega_m$--dependence of the peak counts.}  The probability
distribution function of the convergence \cite{JSW00}, and its
cumulative version, the fractional area of ``hot spots'' on
convergence maps \cite{Wiley+09} are similar to peak counts in the
high-convergence limit and have been shown to have useful cosmology
sensitivity.  The latter statistic is also known as $V_0$, one of the
three Minkowski functionals (MFs) for two dimensional thresholded
fields. MFs are related to peaks and had been proposed as a weak
lensing statistic \cite{Sato_MFs:2001, Guimaraes_MFs:2002}.  More
recently, ref.~\cite{Maturi:2009as} constructed an analytical
approximation to $V_2$, the genus statistic (which also corresponds to
peak counts in the high-threshold limit).  The full set of Minkowski
functionals in the context of weak lensing was studied extensively
both theoretically~\cite{Munshi:2011wu} and in ray-tracing
simulations~\cite{MFs}.  Finally, peak counts have also been studied
in wavelet space~\cite{Pires_WPC:2009}, and found to break the usual
$(\sigma_8, \Omega_m)$-degeneracy that exists between models from the
power spectrum alone.

Preliminary studies \cite{Marian:2008fd, Marian:2009wi} that defined
peaks as local maxima were based on 2D projections of the 3D mass
distribution in low-resolution simulations. WL peak counts with
ray-tracing were subsequently studied by refs.~\cite{DH10, PCWLPC} and
more recently in ref.~\cite{Li2011}.  These were based on simulations
with better mass resolution, and revealed that low--amplitude peaks
(which typically do not correspond to single collapsed dark matter
halos) contain most of the cosmological information.  Various other
aspects of WL peak counts have been further explored. Peaks have been
shown~\cite{Marian:2010mh, Maturi:2011am} to constrain the primordial
non-Gaussianity parameter $f_{\mathrm{NL}}$.  Ref.~\cite{Peaks}
investigated the origin of the cosmologically important low peaks, and
found that they are typically caused by a constellation of 4--8
low-mass halos.  Ref.~\cite{MFs} and \cite{Marian:2011rg} demonstrated
that cosmological constraints from peaks can be tightened by combining
several angular smoothing scales.  WL peak counts have also been
directly compared and found superior to two other commonly used
non-Gaussian statistics, skewness and kurtosis~\cite{Pires_PC:2012}.
Finally, \cite{VanderPlas+2012} study the effect of masked regions on
shear peak counts and show that using Karhunen-Lo\`eve analysis can
mitigate biases on peak count distributions caused by masks, and that
it can reduce the number of noise peaks.

A common limitation to all of the above works is that they are based
on cold dark matter (CDM) simulations (where simulations have been
used), neglecting baryons and all astrophysical processes.  This
leaves an incomplete description of the potential fluctuations and the
corresponding lensing signatures.  This is a particular concern since
previous work has shown that the cooling and condensation of baryons
inside dark matter halos can change the total matter distribution and
has a large impact on the WL power spectrum on small scales
(e.g. refs.~\cite{White,ZhanKnox,Zentner}).  Furthermore, in
astrophysical models that include feedback from active galactic nuclei
(AGN) and supernovae, in addition to cooling and star formation, the
matter distribution can be affected well outside dark matter halos,
modifying the 3D matter power spectrum \cite{vanDaalen+2011}, as well
as the 2D WL power spectrum \cite{Semboloni} out to large scales.

Motivated by these findings, here we quantify the effect of baryons on
the statistics of WL peaks.  Realistic modeling of the astrophysics,
using hydrodynamical simulations, remains challenging, both in terms
of including all of the relevant physical processes correctly, and
also in terms of computational scale.  However, previous studies have
shown that the cooling and condensation of baryons, and the resulting
impact on the total (gas+DM) density profiles of halos, can be modeled
by simple modifications to the halo density profile
\cite{White,ZhanKnox,Rudd+2008,Zentner,Semboloni}.  In particular,
ref.~\cite{Rudd+2008} finds that a simple increase in the
concentration parameter of the universal NFW \cite{UDPHC} profile can
be a good approximation to the results of hydrodynamical simulations,
and can account for the changes in the WL power spectrum
\cite{Zentner}.  We therefore follow this prescription, and manually
steepen the density profile of each individual DM halo identified in
our N-body simulations.

This method is simple to use, and allows us to quantify the effects of
gas cooling. A similar approach could be followed to model the effect
of AGN feedback and other processes, but we leave this to future work.
In this paper, we make a large change to the concentration (increasing
it by $50\%$, compared to the $36\%$ increase that was found to match
simulation results~\cite{Zentner}), thus intentionally amplifying the
impact of baryon cooling on the weak lensing statistics.

The main goal of this paper is to investigate the bias in the
cosmological parameters $w$, $\sigma_8$ and $\Omega_m$ when peak
counts and power spectra are fit neglecting baryonic effects.  Our
results suggest that the bias from the peaks is lower, and also in a
different direction, compared to the power spectrum.  This suggests
that the power spectrum and peak counts can be combined to
``self-calibrate'' WL surveys, i.e.\ to fit cosmological parameters
simultaneously with parameters describing the halo profiles.

The rest of this paper is organized as follows.
In \S~\ref{sec:Methodology}, we describe our calculational procedures,
including the creation of the WL maps through ray-tracing in N-body
simulations, identifying halos and modifying their density profiles,
and creating ``baryonic'' versions of WL maps.  This section also
presents our statistical methodology to compare maps, and our Monte
Carlo procedure of estimating confidence contours and the biases
caused by neglecting the baryonic effects.
In \S~\ref{sec:Results}, we present our results, which include the
effect of baryon cooling on the peak counts and on the power spectrum,
as well as the biases caused by neglecting these effects when fitting
for the three cosmological parameters $w$, $\sigma_8$ and $\Omega_m$.
In \S~\ref{sec: Discussion}, we offer a detailed discussion of our
main results, as well as of several caveats and possible extensions.
Finally, in \S~\ref{sec:conclusions}, we summarize our main
conclusions and the implications of this work.

\section{Methodology}
\label{sec:Methodology}

\subsection{N-body simulations and WL maps}
\label{subsec:simulations}

The cosmological N-body simulations of large-scale structures and
ray-traced weak lensing maps used in this paper are the same as those
in our earlier work \cite{MFs,Peaks}.  We refer the reader to these
publications for a full description of our methodology; here we review
the main features and describe the new features we have implemented to
model the baryonic effects.  We intend to make our data products
publicly available in the future \cite{IG1, IG2}.

A total of 80 CDM-only N-body runs were made with the
\textsc{Inspector Gadget} lensing simulation pipeline \cite{IG1, IG2}
at the New York Blue supercomputer.  NY Blue is part of the New York
Center for Computational Sciences at Brookhaven National
Laboratory/Stony Brook University.

Our suite of 7 cosmological models includes a fiducial model with
parameters \{$\Omega_m=0.26$, $\Omega_\Lambda=0.74$, $w=-1.0$,
$n_s=0.96$, $\sigma_8=0.798$, $H_0=0.72\}$, as well as six other
models.  In each of these six models, we varied one parameter at a
time, keeping all other parameters fixed at their fiducial values; we
thus have WL maps in variants of our fiducial cosmology with
$w=\{-0.8, -1.2\}$, $\sigma_8=\{0.75, 0.85\}$, and $\Omega_m=\{0.23,
0.29\}$.  Note that in the last case, we set $\Omega_\Lambda=\{0.77,
0.71\}$ to keep the universe spatially flat.

To produce the N-body simulations, we first created linear matter
power spectra for the seven different cosmological models with
\textsc{CAMB} \cite{CAMB} for $z=0$, and scaled them back to the
starting redshift of our N-body simulations at $z=100$ following the
linear growth factor.  Using these power spectra to create initial
particle positions, the N-body simulations were run with a modified
version of the public N-body code \textsc{Gadget-2} \cite{Gadget-2}
and its accompanying initial conditions generator \textsc{N-GenIC}.
We modified both codes to allow the dark energy equation of state
parameter to differ from its $\Lambda$CDM value ($w\neq 1$), as well
as to compute WL-related quantities, such as comoving distances to the
observer, at each simulation cube output. Each simulation contains
$512^3$ CDM particles in a cubic box with a side length of
$240h^{-1}{\rm comoving~Mpc}$, allowing a mass resolution of
$7.4\times10^9h^{-1}M_\odot$.  Each of these runs took approximately
1.75 real clock days, using 64 processors on the Blue Gene.

In each of the six non-fiducial cosmological models, we ran 5 strictly
independent N-body simulations (i.e. each with a different realization
of the initial conditions). To minimize the differences between two
cosmologies arising from different random realizations, the initial
conditions for each of those five simulations were matched across the
cosmologies quasi-identically.  This entails recycling the same random
number when drawing mass density modes from the power spectrum for
each cosmology (note that the power spectra themselves of course
differ across the cosmologies).  In the fiducial cosmology, we ran 50
strictly independent simulations -- the first set of 5 to match the
other cosmologies quasi-identically as mentioned above, and an
additional set of 45 to improve the statistical accuracy of the
predictions in the fiducial cosmology (especially the covariance
matrices).  For the ``baryonic'' maps, as described below, only the
first of these 50 simulations was used.  Table \ref{tab:Cosmologies}
lists the sets of simulations in our suite, along with their
cosmological parameters, the number of independent N-body runs, and
the number of pseudo-independent 12 deg$^2$ maps (see below).

\begin{table}
\begin{tabular}{|l|c|c|c|c|c|c|} 
\hline
WL map set & $\sigma_8$ & $w$ & $\Omega_m$ & $\Omega_\Lambda$ & \# of N-& \# of WL\\
  & & & & & body sims & maps \\
\hline
45-sim fiducial & 0.798 & -1.0 & 0.26 & 0.74 & 45 & 1000\\
5-sim fiducial & 0.798 & -1.0 & 0.26 & 0.74 & 5 & 1000\\
$\Omega_m=0.23$ & 0.798 & -1.0 & 0.23 & 0.77 & 5 & 1000\\
$\Omega_m=0.29$ & 0.798 & -1.0 & 0.29 & 0.71& 5 & 1000\\
$w=-1.2$        & 0.798 & -1.2 & 0.26 & 0.74 & 5 & 1000\\
$w=-0.8$        & 0.798 & -0.8 & 0.26 & 0.74 & 5 & 1000\\
$\sigma_8=0.75$ & 0.750 & -1.0 & 0.26 & 0.74 & 5 & 1000\\
$\sigma_8=0.85$ & 0.850 & -1.0 & 0.26 & 0.74 & 5 & 1000\\
\hline
NFW-replaced & 0.798 & -1.0 & 0.26 & 0.74 & 1 & 200\\
``baryonic'' & 0.798 & -1.0 & 0.26 & 0.74 & 1 & 200\\
\hline
\end{tabular}
\caption[]{\textit{Our weak lensing map sets, including the
cosmological parameters, the number of underlying independent N-body
simulations, and the number of weak lensing maps in each
set.}}\label{tab:Cosmologies}
\end{table}

To create the raw WL convergence maps, we used a standard
two-dimensional, flat-sky ray-tracing algorithm closely following
\cite{Hamana:2001vz} with minor modifications as described in detail
in \cite{PCWLPC}. Earlier work with similar algorithms includes
\cite{Schneider:1992, Wambsgaans:1998, Jain:1999ir}. Cubes with
particle positions from the N-body simulations were output every
$80h^{-1}$Mpc in the radial (redshift) direction. While several
independent simulations were used to make each map, boxes from the
same simulations had to be recycled multiple times.  The data cubes at
each redshift were therefore randomly shifted, sliced, and rotated (by
multiples of 90 degrees) to produce pseudo-independent realizations.
The particles were projected onto two-dimensional density planes
perpendicular to the central line of sight of the map. The triangular
shaped cloud (TSC) scheme \cite{Hockney-Eastwood} was used to place
the particles on a grid on these 2D density planes. The Poisson
equation was then solved in Fourier space to convert the surface
density into a gravitational potential. The deflection angles and
convergences were calculated at each plane for each light ray from the
first and second transverse spatial derivative of the 2D gravitational
potential, respectively. Between density planes, the light rays were
assumed to travel in straight lines; $2048\times2048$ light rays were
followed in this fashion for each convergence map.  A total of 1,000
pseudo-independent, 12 deg$^2$ convergence maps were produced in each
CDM-only cosmological model; 200 maps were produced in the sets used
to study the systematic effect of baryons (see below).  The number of
maps in each map set is given in the last column of
Table~\ref{tab:Cosmologies}.

To create the final simulated WL maps, for simplicity, we assumed that
the source galaxies are confined to the single redshift $z_s=2$.
Ellipticity noise from the random orientations of the source galaxies
was added to the convergence maps pixel by pixel, drawing from a
Gaussian distribution corresponding a conservative source galaxy
surface density of $n_{gal}=15$~galaxies/arcmin$^2$, and an
r.m.s. noise in one component of the shear of $\sigma_\gamma=0.22$.
Once noise was added, we smoothed the maps with a finite version of a
$\theta_G=1$arcmin 2D Gaussian filter. This corresponds to the single
most informative and smallest angular scale on which we trust our
maps, based on comparisons with the power spectrum from theoretical
predictions (see~\cite{MFs} for details).  Combining several smoothing
scales would tighten the overall constraints and could change the
biases. We expect that this would not change the qualitative results
in this paper, but we leave an investigation of this issue to future
work.

Peaks are defined as local maxima on the pixelized final mock WL maps,
and are counted in a straightforward fashion, with the convergence in
the central pixel $\kappa_{\rm peak}$ identified as the ``height'' of
each peak. Power spectra are measured numerically from the Fourier
transforms of the same maps following standard techniques.

The differences between cosmological models, and thus the cosmological
parameter dependence of the peak counts and the power spectra, are
computed using pairs of the CDM-only 5-simulation map sets.
Additionally, the covariance matrices of both observables -- peak
counts binned by their height, and power spectrum in bins of $\ell$ --
were computed from the 45-simulation fiducial map set; we have found
that this was necessary for better accuracy (see~\cite{MFs} for
details). Note that we do not consider the dependence of the
covariance matrices themselves on cosmology (in principle, this
dependence could help improve constraints).

Finally, in order to study the impact of baryons, we create two more
sets of WL maps.  These are based on the fiducial CDM model, and a
single N-body simulation.  In this simulation, we identify all the DM
halos in the 3D simulation cubes, and replace them with spherically
symmetric halos with analytic NFW \cite{UDPHC} density profiles.  The
ray-tracing procedure is then repeated exactly as before, and a new
set of 200 ``NFW-replaced'' maps is created.  This set is used to
predict the expectation values of the peak counts and the power
spectra in the fiducial cosmology, in the absence of baryons.
Finally, beginning with the same 3D data, the concentration parameter
``$c$'' of each spherical NFW halo is increased by 50\%, and the
ray-tracing and map-making procedure is once again repeated, to create
a corresponding set of 200 ``baryonic'' maps.  This last set is used
to find the expectation values of the peak counts and the power
spectra in the fiducial cosmology, in the presence of baryon cooling.
This procedure guarantees that any differences between an individual
``NFW-replaced'' map and the corresponding ``baryonic'' map is caused
only by the changes in the halo concentration (and not from the halo
replacements or from having different random realizations).  Note that
ideally we would want 1,000 pseudo-independent realizations of the
baryonic maps, to match the number we have for the CDM-only map sets.
However, to keep computational costs feasible, we have produced only
200 baryonic maps from a single N-body run.  Nevertheless, we
replicate each of these 200 baryonic maps 5 times, each time adding a
different random noise realization. This provides a better statistical
sampling of the noise (in particular, a more accurate determination of
the average effect of the noise).

\subsection{Identifying and replacing dark matter halos}
\label{subsec:Haloreplacement}

In this section we describe in detail how we identify, remove, and
re-insert halos into the N-body simulations.

Once the N-body simulations are generated, we use the publicly
available AMIGA halo finder(\citep{AHF}, hereafter AHF) to identify
collapsed halos in our N-body runs. Its output consists of the 3D
positions and the tagged set of particles belonging to each halo, the
central position of the halo (defined as the local maxima of the
density field), as well as the total number of halo particles inside
spherical shells at several different discrete radii, from which the
spherically-averaged density profiles can be calculated.  Depending on
the redshift, AHF identifies a total of $2.5\times10^5-2.6\times10^5$
halos in our box, with masses in the range $1.5\times
10^{11}h^{-1}M_\odot-5\times10^{14}h^{-1}M_\odot$.  We then remove all
the particles belonging to all of the halos from the N-body
simulations, and add back the density profiles fitted to the
identified halos, assuming halos are spherically symmetric and
described by NFW profiles.  The details of the fitting will be given
in \S\ref{subsec:haloprofile} below.

In the procedure of replacing halos, three non-trivial points
regarding mass conservation need to be clarified.  The first point
concerns sub-halos.  When a parent halo and a sub-halo share a common
structure, the halo finder saves the shared particles in both the
parent and the sub-halo profiles.  As a result, in our procedure, the
particles within halos are removed one time, but the shared parts of
halos are added back repeatedly.  We have found that this can cause an
artificial $5\%$ increase in total (halo plus subhalo) mass.  To avoid
this problem, and to conserve mass, we sort the halo catalog by
mass. Beginning with the lowest-mass halo, we consider halos in this
ranked list one-by-one, always comparing the position of the center of
a halo with the positions of all of the halos further down the list
(with higher masses).  If the center of a halo is found to fall inside
the virial radius of any of the higher-mass halos, we subtract the
mass of the smaller halo from that of the larger one.  When we
re-insert the larger halo into the 3D simulation box, we use the
analytical NFW profile with the reduced mass.

The second point concerns discretization of the analytic NFW
profiles. In our map-making procedure, the projected halo density
profiles are evaluated only at the discrete grid points on our 2D
density plane.  As a result, each halo effectively contributes a total
mass given by a ``2D trapezoidal integral'', rather than its actual
(analytically calculable) mass.  In principle, the actual halo
profiles are known, and the projected mass could be resolved to
arbitrary accuracy; in practice, we have found this to be
computationally too expensive. Instead, in order to conserve the
total mass, we normalize the discretized surface density profile by
multiplying by the ratio of the actual mass of the halo and the total
projected mass of the discretized profile.

Finally, a third point concerns halos that are ``truncated'', either
in the transverse direction (because they are too close to the edge of
the 12 deg$^2$ map) or in the redshift direction (because they are too
close to the back or front side of the underlying 3D simulation slice;
note that the raw 3D simulation cubes have periodic boundary
conditions, but the slices used to make the maps do not~\cite{MFs}).
In the former case, we add the projected halo density profiles as we
do for the normal halos, applying periodic boundary conditions.  In
the latter case, the surface density of a truncated halo is taken by
summing the discretized density profile over the region inside the
simulation box.  We do not re-normalize the profiles of these
truncated halos in the redshift direction, because it is difficult to
determine how much mass was actually lost in the truncation. However,
we have shown in our previous study~\cite{Peaks} that the effect of
edge halos can be safely ignored for studying the peak counts.

After accounting for the sub-halos and normalizing the discretized
profiles, the fractional difference between the total mass removed and
the total mass added back is less than $0.2\%$.  

\begin{figure}[htp]
\centering
\includegraphics[width=8 cm]{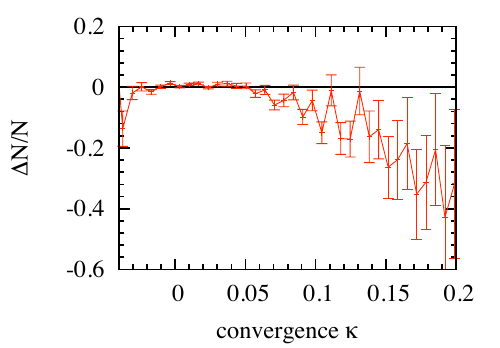}\\ 
\includegraphics[width=8 cm]{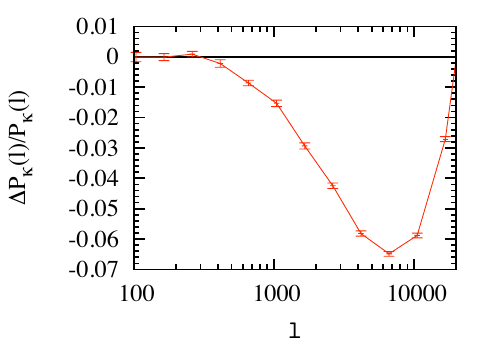}\\
\hfill
\caption[]{\textit{The effect of replacing halos in our N-body
simulations by spherically symmetric NFW halos. In the upper panel, we
show the fractional difference $\langle\Delta N\rangle/\langle
N\rangle$ in the peak counts caused by this halo replacement, in
convergence bins of width $\Delta\kappa=0.00675$. In the lower panel,
we show the corresponding fractional difference in the power spectrum
$\langle\Delta P_\kappa\rangle/\langle P_\kappa\rangle$, in
logarithmic bins of the spherical harmonic index with a width
$\Delta\log\ell=0.2$.  Data points are averaged over 100
realizations. The error bars in the top panel are estimated as the
standard deviation of $\Delta N$, divided by $\langle N\rangle$ (and
similarly as the standard deviation of $\Delta P_\kappa$, divided by
$\langle P_\kappa\rangle$, for the power spectrum).  The source
galaxies are assumed to be at redshift $z_s=2$, and the convergence
maps have been smoothed with a 1 arcmin Gaussian filter.  }}
\label{fig:ModelCheck}
\end{figure}

To check the ultimate accuracy of the halo replacement procedure, in
Figure~\ref{fig:ModelCheck}, we plot the fractional difference in the
peak counts and the power spectrum caused by the halo replacement in
the fiducial cosmology.  The data points are averaged over 100
realizations.  For this check, we used the noiseless versions of the
WL maps.  As the upper panel shows, the halo replacement works very
accurately for peaks below $\kappa\lesssim 0.06$.  The difference is
still smaller than $10\%$ for peaks below $\kappa\lesssim 0.1$,
however there is a systematic decrease in the number of peaks that
becomes significant for higher amplitude peaks.  This decrease is
caused by our treatment of the sub-halos.  High peaks are typically
caused by a single massive halo. Massive halos contain many sub-halos,
and because of the substantial mass loss due to tidal stripping,
sub-halos are found preferentially away from the center of the host
halos~\citep{NK2005}.  Recall that when we add back an NFW parent
halo, in order to conserve mass, we subtract the total mass in
subhalos.  When we perform this subtraction, we assume, for
simplicity, that the mass in the subhalos is distributed radially in a
smooth fashion, following the density profile of the parent halo.
This procedure therefore over-subtracts mass near the halo center, and
under-subtracts mass near the halo outskirts.  The overall effect of
our procedure is to decrease the central density (by a few percent).
Since the line of sight to a high peak typically passes very close to
the center of the corresponding massive halo, this diminishes the
height of the peak, and results in fewer high peaks.  A similar
explanation holds in the case of the power spectrum, shown in the
lower panel.  The halo replacement works very accurately for large
scales, with $\Delta P/P \lesssim 2\%$ for $l < 1000$.  But at $l\gsim
5,000$, where the one-halo term dominates the power spectrum, there is
a significant ($\sim 6\%$) decrease in power.

We emphasize that the inaccuracies due to halo replacement are
quantified here only as a reassurance that we did not, in the process,
gravely modify the WL observables.  We are only interested in the
impact of baryons; indeed, the impact of baryon cooling in the real 3D
halos should be similar to the increase in concentration parameters
for the spherically symmetric halos.

\subsection{Fitting halo density profiles}
\label{subsec:haloprofile}

In this section, we describe our procedure for the fitting the halo
density profiles and quantify how well this procedure works.

We assume all halos and subhalos follow a universal NFW profile
\cite{UDPHC},
\BA\label{eq:NFW1}
\rho_{\rm nfw}(r)&=&\frac{\rho_s}{(r/r_s)[1+(r/r_s)]^2}
\EA 
where $r$ is the radius from the halo center, and $r_s$ and $\rho_s$
are a characteristic radius and density. The profile is truncated at
$R_{180}$, inside which the mean overdensity is 180 times the matter
density of the universe at redshift $z$.  This convention is
consistent with the settings in the AHF halo finder.  The
concentration parameter is given by $c_{180}\equiv R_{180}/r_s$.

The NFW density profile is uniquely defined by the concentration
parameter $c_{180}$, and the normalization factor $\rho_s$. To obtain
$c_{180}$, we chose to fit the normalized cumulative density profile
$F(x)$, i.e. the fraction of mass within a certain radius $r$, as a
function of $x=r/R_{180}$,
\BA\label{eq:NFW2}
F(x)&=&\frac{\log(c_{180}x+1)-c_{180}x/(1+c_{180}x)}{\log(c_{180}+1)-c_{180}/(1+c_{180})}.
\EA 
Here $R_{180}$ is taken as the actual radius of halos found by the
halo finder.  We fit $F(x)$ to find $c_{180}$ and calculate the
normalization factor $\rho_s$ as:
\BA\label{eq:NFW3}
\rho_s&=&\frac{M_{180}}{4\pi r_s^3 (\log(c_{180}+1)-c_{180}/(1+c_{180}))}.
\EA 
Here $M_{180}$ is the actual halo mass returned by the halo finder.
For some halos, we find that $c_{180}$ does not converge to an
acceptable number ($1.1 \le c_{180} \le 50$), or the halo mass is too
low and the halo finder cannot compute a radial profile. We
distinguish these as unfitted halos, and for these halos we use an
analytical formula to obtain $c_{180}$, adapted from Eq.~(5) in
ref.~\citep{Concentration}:
\BA\label{eq:NFW4}
c_{200}&=&4.67~(M_{200}/10^{14}h^{-1}{\rm M_\odot})^{-0.11}/(1+z).
\EA 
Eq.~(\ref{eq:NFW4}) adopts a different convention from ours -- the
profile is truncated at the radius inside which the mean overdensity
is 200 times the critical density of the universe.  We therefore
extend the NFW profile for a halo with mass $M_{200}$ outward to a
radius where the mean interior density is 180 times the background
matter density.  This extrapolation yields a relation between the
input $M_{180}$ and $M_{200}$ that depends on $c_{180}$; using
Eq.~(\ref{eq:NFW4}) and the definitions of the various quantities above,
the parameter $c_{180}=R_{180}/r_s$ can be found iteratively.

The unfitted halos account for $\sim 2\%$ of all halos in our
simulations.  This is unsurprising, since, for example, halos that are
undergoing major mergers, or have not yet relaxed from a recent major
merger, have no reason to follow NFW profiles \citep{Concentration}.

\begin{figure}[htp]
\centering
\includegraphics[width=8 cm]{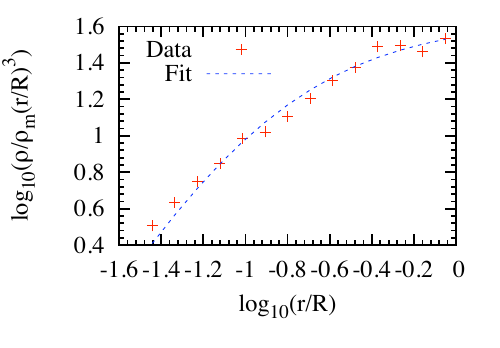}\\ 
\hfill
\caption[]{\textit{The figure shows the spherically averaged density
profile $\rho(r)r^3$ (red crosses), for a halo with a mass of
$8.7\times 10^{13}h^{-1} M_\odot $ at redshift $z=0$, along with the
corresponding best-fit NFW profile (dashed blue curve).  The quality
of the fit shown in this figure is typical of halos in our
simulations.  Here $\rho_m$ is the mean matter density of the universe
and $R_{180}$ is (approximately) the virial radius of the halo.  The
best-fit NFW profile was obtained by a least-squares fit to the
normalized cumulative density profile (Eq.~(\ref{eq:NFW2})). }}
\label{fig:fitex}
\end{figure}

We find that the rest of our halos follow the NFW profiles accurately.
In Figure~\ref{fig:fitex}, we show the spherically averaged density
profile $\rho(r)r^3$, for a halo with a mass of $8.7\times
10^{13}h^{-1} M_\odot $ at redshift $z=0$, along with the
corresponding best-fit NFW profile.  The quality of the fit shown in
this figure is typical of halos in our simulations.

\begin{figure}[htp]
\centering
\includegraphics[width=8 cm]{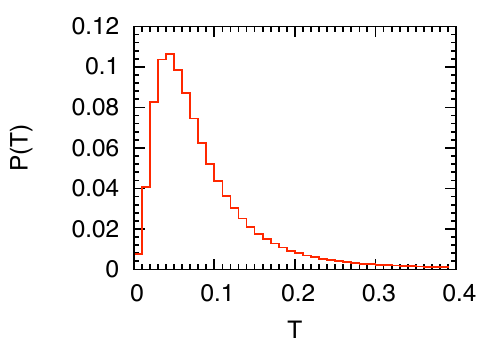}\\ 
\hfill
\caption[]{\textit{The distribution of the statistic $T$, defined as
the fractional deviation between the fitted and the actual normalized
cumulative density profile (Eq.~(\ref{eq:fit1})), averaged over radii,
for the halos in our N-body simulations.  The figure demonstrates that
typically, the NFW profiles are accurate to within 5\%, although there
is a tail of outliers extending to poorer fits.  Approximately $2\%$
of our simulated halos could not be fit by NFW profiles at all (and
are not shown in this figure).  }}
\label{fig:fitstatistics}
\end{figure}

As a quantitative test of the accuracy of our fitting, we define a
statistic
\BA\label{eq:fit1}
T&=&\frac{1}{n}\sum^{n}_{i=1}\left|\frac{F_i-F(x_i)}{F_i}\right|.
\EA 
Here $n$ is the number of data points (i.e. in radius) for one halo,
$F_i$ and $F(x_i)$ is the $i$th data point and the fitted value at
that radius.  We plot the distribution of T in
Figure~\ref{fig:fitstatistics}. The figure demonstrates that
typically, the NFW profiles are accurate to within 5\%, although there
is a tail of outliers extending to poorer fits.  As mentioned above,
approximately $2\%$ of our simulated halos could not be fit by NFW
profiles at all (and are not shown in this figure).

\subsection{Statistics \protect\footnote{This discussion closely follows ref.~\cite{MFs}.}}

We refer to the different statistics one can obtain from a 2D WL
map---e.g. power spectrum, peak counts, etc.---as $N_i$.  The index
$i$ in the components $N_i$ of the vector $\mathbf{N}$ labels
different heights for the peaks, and different multipoles for the
power spectrum. We divide the peak counts into 30 bins in height
$\kappa_{\rm peak}$ in the range $0.023\leq \kappa_{\rm peak}\leq
0.08$ and, we use a second set of 30 bins above $\kappa_{\rm peak}\geq
0.08$.  Similarly, we divide the power spectrum into 30 multipole bins
$\ell$ in the range $100 \leq \ell \leq 2\times10^4$.  For
computational reasons, the power spectrum is first pre-computed and
stored for 1,000 ``pre-bins'' spaced linearly in multipole $100 \leq
\ell \leq 1\times10^5$ and is only later combined into the 30 larger
bins used for the final computation.

To constrain cosmology, we are interested in the true ensemble
average, over an infinite number of realizations within a single
cosmology (hereafter denoted by brackets $\langle\ \rangle$) as a
function of cosmological parameters $\mathbf{p}=\{\Omega_m, w,
\sigma_8\}$, as well as the ensemble covariance.\footnote{More
  generally, one could utilize the full probability distribution of a
  statistic, not just its average, as well as the cosmology-dependence
  of all higher-order correlations, not just the covariance; we will
  not investigate these issues in the present paper.}  Since these are
not available, we estimate them from a finite number of our
simulations.  Averaging over the pseudo-independent map realizations
within a given cosmology, we can obtain an estimate for the ensemble
average by
\BE\label{SM}
\langle{N}_i(\mathbf{p})\rangle\approx\overline{N}_i(\mathbf{p})\equiv\frac{1}{R}\sum_{r=1}^R
N_i(r, \mathbf{p}),
\EE
where $N_i(r, \mathbf{p})$ is the statistics vector measured in a
single map and $r$ runs over our $R=1,000$ maps.  We call this
estimate the \textit{simulation mean}.  It differs from the true
ensemble average both because of the limited number of realizations
and also because of the limitations inherent in our simulations, such
as limited resolution.  In the absence of a fitting formula for the
peak counts in the non-Gaussian case (analogous to the power spectrum
formula from \cite{Smith+03} in the nonlinear regime) the simulation
mean serves as our proxy for theoretically predicted peak
counts.\footnote{The ensemble average for the peak counts can be
computed exactly for a Gaussian random field~\cite{SCBRF}; however,
this is a poor approximation to the lensing peaks~\cite{Peaks}.}

We can form this estimate only for the 7 selected cosmologies where we
have run simulations (Table~\ref{tab:Cosmologies}). For cosmologies
with other parameter combinations $\mathbf{p}$, we have to interpolate
(and in a few cases extrapolate) between these 7 points in the 3D
parameter space.  We thus construct a first-order Taylor expansion
around our fiducial cosmology:
\BA\label{Taylor}
\hspace{-0.4cm}\overline{N}_i(\mathbf{p})&\approx&\overline{N}_i(\mathbf{p^{(0;NFW)}})+\nonumber\\
& &\hspace{-1.3cm}+\sum_\alpha \frac{\overline{N}_i(\mathbf{p^{(\alpha;CDM)}})-
\overline{N}_i(\mathbf{p^{(0;CDM)}})}
{p^{(\alpha)}_\alpha-p^{(0)}_{\alpha}}
\cdot(p_\alpha-p^{(0)}_{\alpha}).
\EA
The index $\alpha=1,2,3$ refers to either $\Omega_m, w$, or
$\sigma_8$, and the sum counts through all 3 parameters.  The fraction
on the right-hand-side of Eq.~(\ref{Taylor}) is a finite difference
derivative along the direction of the parameter $\alpha$, computed
using the fiducial cosmology ($\mathbf{p^{(0)}}$) and a cosmology in
which the parameter $p_\alpha$ was changed from the fiducial case
($\mathbf{p}^{(\alpha)}$).  As mentioned above, the 5-simulation CDM
map sets with quasi-identical initial conditions were paired and used
for the computation of this derivative, but the simulation mean for
the NFW-replaced set was used as the expansion point of the Taylor
series (the first term on the RHS). We indicate this explicitly in
Eq.~(\ref{Taylor}) by the superscripts ``CDM'' and ``NFW''.

If this non-fiducial cosmology is chosen such that
$p^{(\alpha)}_\alpha-p^{(0)}_{\alpha}$ is positive for all three
parameters, we call it a ``forward derivative'', if it is negative, we
call it a ``backward derivative''. We will also refer below to a
``2-sided derivative'', which switches automatically between the
forward and backward cases as needed for each parameter, corresponding
to the octant in the 3D parameter space where the interpolation is
performed. For simplicity, we use the backward derivative throughout
this paper unless explicitly noted otherwise (and below we highlight
some important differences in our results obtained with other
derivatives).

Similarly to the simulation mean, we estimate the ensemble covariance
matrix from the simulations, ${\rm Cov}( N_i, N_j) \approx C_{ij}$,
where
\BE
\label{eq:covariance matrix}
C_{ij}(\mathbf{p})\equiv\frac{1}{R-1}\sum_{r=1}^R [N_i(r,\mathbf{p})-\overline{N}_i(\mathbf{p})][N_j(r,\mathbf{p})-\overline{N}_j(\mathbf{p})].
\EE
This covariance matrix contains contributions both from the sample
variance of the signal and from the galaxy shape noise.  In the case
of the power spectrum, the Gaussian random noise in different $\ell$
bins should be uncorrelated, and therefore contribute only diagonal
terms (in practice, the cross-terms are small but nonzero in our
finite set of realizations).  However, adding noise to the maps has a
nonlinear effect on peak counts, and introduces off-diagonal terms, as
well.  As mentioned above, in practice, we only evaluate and use the
covariance matrix in our fiducial 45-simulation map set.

\subsection{Monte Carlo parameter estimation contours}
\label{sec:Monte Carlo}

Each of our WL maps spans a 12 deg$^2$ field of view, yet we wish to
obtain parameter contours for a 20,000 deg$^2$ full-sky survey, such
as LSST.  Thus we employ a parametric bootstrapping approach to
generate approximations to full-sky maps. In this procedure, we
randomly select a map from our set of 1,000 maps, with replacement,
$20,000/12\approx1,667$ times.  The effective solid angle of this
larger ``composite map'' is then 20,000 deg$^2$, as desired.  The map,
of course, is not a true composite -- we merely compute the
observables for each of the 1,667 patches individually, and then add
or average over them to get their values for the full-sky map. We
create 10,000 such full-sky maps to obtain smooth parameter contours
in our Monte Carlo procedure.  While this process cannot generate
information not present in the 1,000-map set, the bootstrap procedure
has a large benefit. Individual 12 deg$^2$ maps constrain parameters
only poorly, and large fluctuations in $N_i$ require parameter
extrapolations far outside the range of our simulations.  The
averaging in the bootstrap procedure suppresses these fluctuations, so
the $N_i$ are similar to those from the simulations.

To estimate the cosmological parameter error contours from the set of
baryonic maps, we use $\chi^2$-minimization to find the best-fit
cosmological parameter combination for each of the 10,000 bootstrapped
full-sky maps.  Here $\chi^2$ is defined by
\BE 
\label{chi2}
\chi^2(r_b, \mathbf{p})\equiv\sum_{i,j}\, \Delta N_i(r_b, \mathbf{p})\,[{\rm Cov}^{-1}\mathbf{(p^{(0)})}]_{ij}\, \Delta N_j(r_b,\mathbf{p})
\EE
where
\BE
\Delta N_i(r_b,\mathbf{p})\equiv N_i(r_b,\mathbf{p^{(0;baryon)}})- \langle{N}_i(\mathbf{p})\rangle.
\EE
Note that $r_b$ here counts the full-sky maps and therefore runs over
$r_b=1,\dots,10,000$. For each Monte Carlo realization $r_b$, we
minimize $\chi^2$ with respect to $\mathbf{p}$ using a simulated
annealing algorithm. This is a popular technique based on Markov Chain
Monte Carlo with a decreasing ``temperature'', which helps the
algorithm to get out of local minima (we had success with this
algorithm in our earlier work~\cite{MFs}).  We emphasize that the
matrix Cov in Eq.~(\ref{chi2}) can be reasonably arbitrary and does
not even have to be the covariance matrix; this makes this method
robust with respect to noise in the covariance matrix
estimate~\cite{MFs}.

The bootstrapping procedure explained above returns 10,000 sets of
best-fit parameters for each full-sky realization; the distribution of
these best-fit points can be used to draw confidence contours at
desired confidence levels (68\% in this paper, unless stated
otherwise) based on the density of the best-fit points.

We indeed find that the best-fit points lie within the parameter range
of our simulations, so that interpolation can be used during the
best-fit procedure. The only exception is the high-$\ell$ power
spectrum, which required a modest extrapolation in the parameter
space.  In contrast, the scatter of best-fit points from the original
small 12 deg$^2$ maps was so large that a significant fraction of the
best-fits lie outside the simulated region, requiring large
extrapolations.  Bootstrapping also eliminates the need to re-scale
the contours to correspond to a full-sky survey.

\section{Results}
\label{sec:Results}

\subsection{Effect of baryons on the statistics}
\label{subsec:EffectofBaryons}

\begin{figure}[htp]
\centering
\includegraphics[width=8 cm]{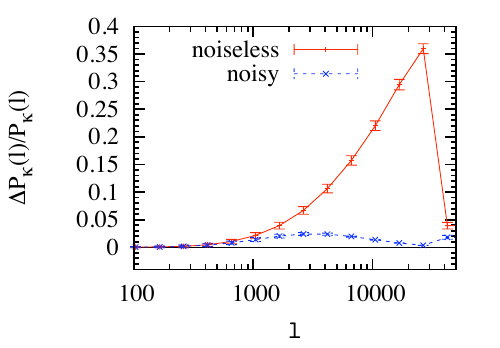}\\ 
\hfill
\caption[]{\textit{The fractional difference $\langle\Delta
    P_\kappa\rangle/\langle P_\kappa\rangle$ in the power spectrum
    caused by the 50\% boost in the concentration parameter of each
    NFW halo, shown in models without (red solid curve) and with
    galaxy shape noise (blue dashed curve; the noise corresponds to a
    source galaxy surface density of $n_{gal}=15~{\rm arcmin}^{-2}$ at
    redshift $z_s=2$). The data points are averaged over 200
    realizations in logarithmic bins of width
    $\Delta\log\ell=0.2$. Error bars are estimated as the standard
    deviation of the $\Delta P_{\kappa}$, divided by $\langle
    P_\kappa\rangle$.  We have artificially increased the error bars
    by a factor of 10 for visual clarity. The smallness of the error
    bars indicate that the impact of baryons on the power spectrum is
    highly systematic, i.e. very similar in each realization.}}
\label{fig:DiffPs}
\end{figure}

\begin{figure}[htp]
\centering
\includegraphics[width=8 cm]{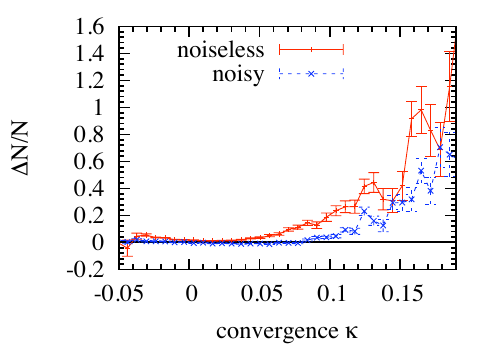}\\ 
\hfill
\caption[]{\textit{Similar to Figure~\ref{fig:DiffPs}, except we show
    the fractional difference in the peak counts, $\langle\Delta
    N_{\rm peak}\rangle/\langle N_{\rm peak}\rangle$, caused by the
    same 50\% boost in the halo concentration parameters.  The data
    points are averaged over 200 realizations, and shown in
    convergence bins of width $\Delta\kappa=0.00675$.  The error bars
    are shown at their original size. Compared to the power spectrum,
    the impact of baryons on the peak counts is less systematic,
    i.e. varies more significantly between realizations.}}
\label{fig:DiffHisto}
\end{figure}

We begin the presentation of our results with the effect of baryons on
the statistics themselves.  In Figure~\ref{fig:DiffPs}, we show the
fractional difference $\langle\Delta P_\kappa\rangle/\langle
P_\kappa\rangle$ in the power spectrum, caused by the 50\% boost in
the concentration parameter of each NFW halo, shown in model with and
without galaxy shape noise.  The data points are averaged over 200
realizations in logarithmic bins of width $\Delta\log\ell=0.2$.  As
the figure shows, in the noiseless case, a 50\% increase in
concentration causes a strong increase in the small scale power, by
about $20\%$ at $\ell=10^4$.  The effect is much smaller on large
scales ($\ell\lsim2000$).  This agrees with the qualitative
conclusions in ref.~\cite{Zentner}, who used a similar toy model.  The
drop beyond $\ell\gsim 3\times10^4$ is due to our 1-arcmin smoothing.
In the noisy case, the fractional difference has a turnover at the
much larger scale of $l\approx3,000$; this is the scale at which the
galaxy shape noise starts to dominate.  Note that the absolute
difference $\langle\Delta P_\kappa\rangle$ in the power spectrum in
the noisy case is exactly the same as in the noiseless case. Therefore
even the small $<2\%$ fractional difference in the noisy case can cause a
strong bias in inferred cosmology when the baryonic power spectrum is
fitted by the non-baryonic model (see detailed discussion below).
 
In Figure~\ref{fig:DiffHisto}, we show the fractional difference in
the peak counts, $\langle\Delta N_{\rm peak}\rangle/\langle N_{\rm
  peak}\rangle$, caused by the same artificial 50\% boost in the halo
concentration parameter.  The data points are again averaged over our
200 maps, and shown in convergence bins of width
$\Delta\kappa=0.00675$. As the figure shows, in both the noiseless and
the noisy case, there is a strong increase in the number of high
peaks, but very little change in the number of low peaks.  The reasons
why the low peaks are robust to the change in the concentration
parameter will be discussed below.

\subsection{Cosmological constraints and biases}
\label{subsec:InferredBias}

We next present the biases in the inferred best-fit cosmological
parameters when the baryonic effects are ignored. We define the
``low'' and ``high'' peaks to have heights $\sigma_{noise}\leq
\kappa_{\rm peak}\leq 0.08$ and $\kappa_{\rm peak}\geq 0.08$, where
$\sigma_{\rm noise}=0.023$ is the r.m.s. of $\kappa$ from galaxy shape
noise.  We similarly define the power spectrum on ``very large'',
``large'', and ``small'' angular scales (or ``very low'', ``low'', and
``high'' $\ell$ ranges) to be those with $100\leq\ell\leq1000$,
$100\leq\ell\leq2000$ and $2000\leq \ell \leq 2\times10^4$,
respectively. Note that two of these ranges overlap partially.

As described in \S~\ref{sec:Monte Carlo}, we generate the distribution
of best-fit points in the 3-dimensional parameter space of $\sigma_8$,
$w$, and $\Omega_m$.  Projecting this 3D distribution in one of the 3
dimensions can then be used to define the $68\%$ joint confidence
contours on the remaining two parameters, marginalized over the third.
As also described above, the baryonic model was compared to the
non-baryonic NFW-replaced model, linearly interpolated with the
backward derivative, using noisy maps with $z_s=2$ and 1 arcmin
smoothing.  We apply 30 nearly logarithmic bins to the whole range
$100\leq\ell\leq2\times10^4$ of the power spectrum.  Note that this
leaves fewer than 30 bins in each of the very low, low and high-$\ell$
ranges as defined above (namely, 9, 17, and 13 bins, respectively) The
nearly logarithmic bins are generated by choosing the bin boundaries
so that the 1,000 pre-bins are assigned to 30 bins with nearly equal
logarithmic spacing. For very low $\ell$, this results in bins that
are linearly spaced, because there are not enough pre-bins to achieve
truly logarithmic spacing.  For the peaks, the bin boundaries are
chosen so that each of the 30 bins for the low peaks contains the same
number of peaks, and likewise we use 30 equal-count bins for the high
peaks. This split in the equal-count procedure avoids having too few
bins for the high peaks.

\begin{figure*}[htp]
\centering
\includegraphics[width=16 cm]{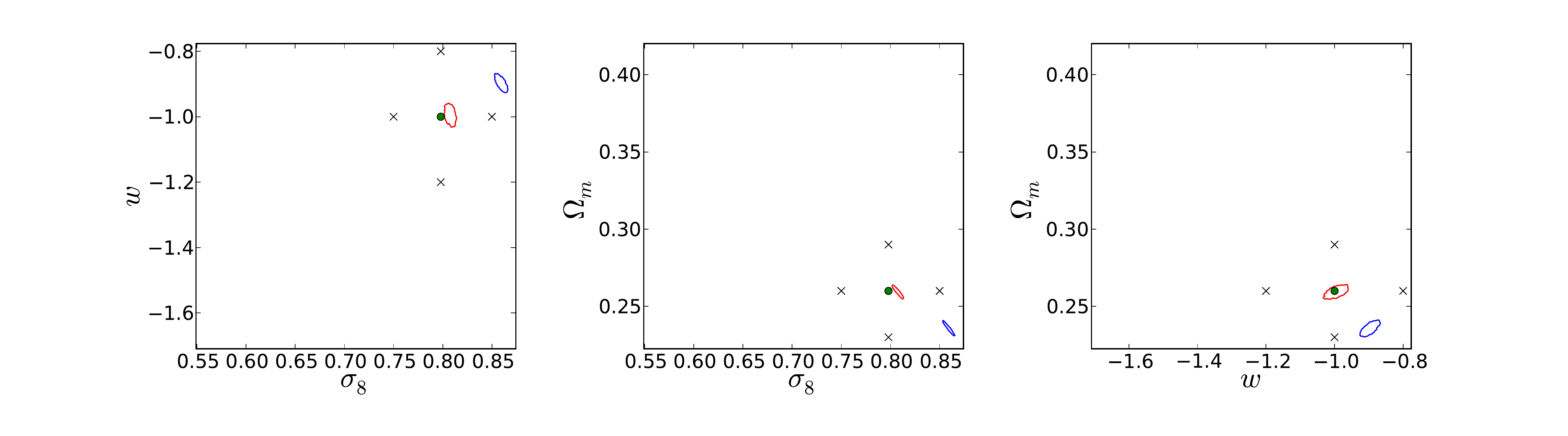} \\
\includegraphics[width=16 cm]{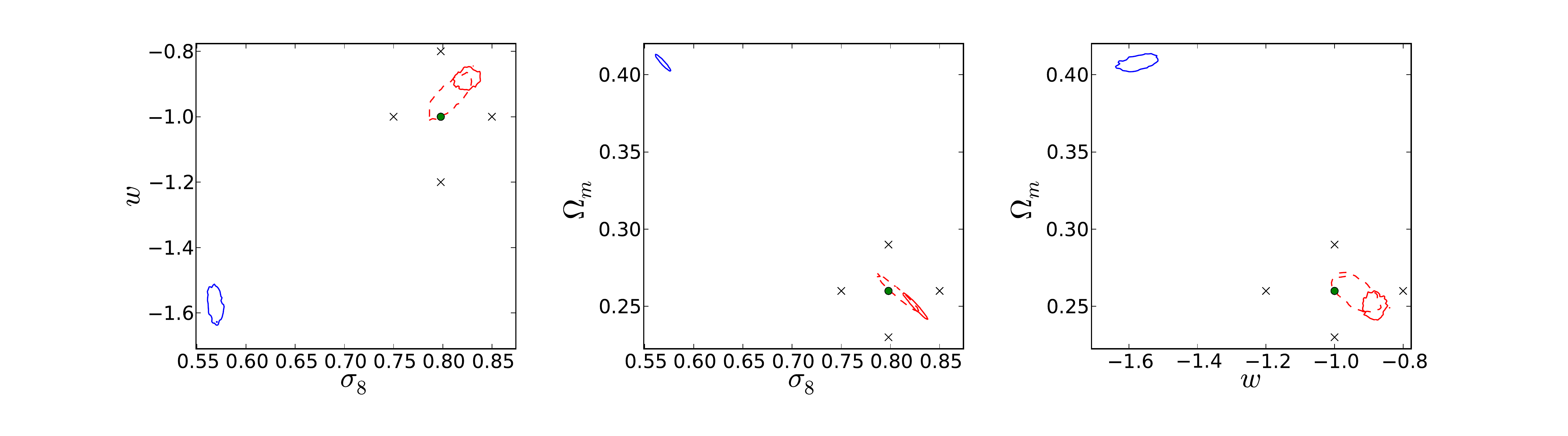}
\hfill
\caption[]{\textit{The joint 68\% confidence contours on pairs of
    cosmological parameters (marginalized over the third), obtained
    from peak counts and the convergence power spectrum, in a full sky
    ($2\times10^4 deg^2$) survey.  The location of the correct
    fiducial cosmology is marked by a green dot in each panel. For
    reference, crosses mark the other cosmologies we simulated. 10,000
    Monte-Carlo realizations of the baryonic model were fit by models
    neglecting the baryon effects, which can produce a bias in the
    inferred parameters.  In the upper panels, results from the low
    ($\sigma_{noise}\leq \kappa_{\rm peak}\leq 0.08$) and high
    ($\kappa_{\rm peak}\geq 0.08$) peaks are shown by solid red and
    solid blue contours. In the lower panels, results from the
    very-low ($\ell \leq1000$), low ($\ell \leq 2000$), and
    high-$\ell$ ($2,000\leq \ell \leq 2\times10^4$) power spectra are
    shown by dashed red, solid red and solid blue contours,
    respectively.  The bias is strongest from the small-scale
    (high-$\ell$) power spectrum; it is nearly negligible for the low
    peaks and for the very-low $\ell$ power spectrum.}}
\label{fig:Inferred bias}
\end{figure*}

In Figure~\ref{fig:Inferred bias}, we show the $68\%$ confidence
contours obtained from the peak counts, along with those from the
power spectra, for a 2$\times10^4$-deg$^2$ full-sky survey.  We have
approximately 2,000 low peaks and 250 high peaks in every 12 deg$^2$
field.  The offsets between the center of the underlying 3D
distribution (i.e. the best-fit model in 3D) and the correct fiducial
values are listed in Table~\ref{tab:Inferred bias}, and correspond to
the biases on the inferred cosmological parameters.  The contours in
the 2D plots of Figure~\ref{fig:Inferred bias} are always marginalized
over the third parameter that is not displayed, i.e. they are
projections of the full 3D best-fit-point distribution onto the 2D
plane, and the contours are calculated in this projected space.
Recall that we only have 1000 (pseudo-)independent 12 deg$^2$ maps,
from which 1,666 maps are drawn randomly with replacement in the
bootstrapping procedure.  As bootstrapping cannot improve the accuracy
of the mean, it cannot improve the determination of the locations of
the centers of the contours, and affects only their size. The
``1$\sigma$'' error of the centroids (in these 2D planes) should
therefore be larger than the size of the contours (the 68\% confidence
levels from the LSST-like survey) by a factor of approximately
$\sqrt{1666/1000}\approx 1.3$.  This is in the limit that random noise
dominates the dispersion in the best-fit locations.  In the opposite
limit, i.e. if the dispersion is dominated by the variations among the
underlying 200 baryonic maps, then the peak likelihood locations in 2D
would be larger by an additional factor of $\sqrt{5}$.

From Figure~\ref{fig:Inferred bias}, we can roughly estimate that the
joint 2D errors from the peak counts extend over
$\delta\sigma_8\approx 0.005$, $\delta w\approx 0.03$ and $\delta
\Omega_m\approx 0.005$; the corresponding uncertainties in the
locations of the best-fits in 2D are therefore
$\delta\sigma_8^\prime\approx 0.0065$, $\delta w^\prime\approx 0.04$
and $\delta\Omega_m^\prime\approx 0.0065$.  A similar estimate can be
done for the power spectrum.  Given these uncertainties, we conclude
that the low peaks and the very low-$\ell$ power spectrum are both
robust to baryonic cooling, with the inferred biases consistent with
zero; the exception is that low peaks have a small bias in $\sigma_8$.
The high peaks and the high-$\ell$ power spectrum both have
significant biases, $\Delta\sigma_8\approx 0.061$, $\Delta w\approx
0.10$ and $\Delta\Omega_m\approx -0.024$ from peak counts, and
$\Delta\sigma_8\approx -0.23$, $\Delta w\approx -0.58$ and
$\Delta\Omega_m\approx 0.15$ from the power spectrum (see
Table~\ref{tab:Inferred bias}).  The low bias, at least within the
context of our model, and the high information content of the low
peaks suggests the value of including them in cosmological analyses
for dark energy (see further discussion in \S~\ref{subsec:Constraint}
below).

Perhaps our most interesting result is that the biases from the peaks
and the power spectra are in different directions in the 3D parameter
space. The bias from the high peaks, in particular, is in a direction
nearly opposite from the (much stronger) bias from the high--$\ell$
power spectrum; they are also in different directions from the
low--$\ell$ power spectrum.  This opens up the possibility for
self-calibration, whereby baryonic parameters, such as the halo
concentration parameter in our case, could be determined
simultaneously with the cosmological parameters. Only for the correct
concentration parameter will the contours from the different
observables align.

\begin{table}
\begin{tabular}{|c|ccc|} 
\hline
Observable  & $\Delta\sigma_8$ & $\Delta w$ & $\Delta \Omega_m$\\
 \hline
  Low Peaks & 0.009 & 0.01 & -0.001 \\
 \hline
 High Peaks & 0.061 & 0.10 & -0.024 \\ 
  \hline
 Low-$\ell$ Pow. Spec. & 0.03 & 0.11 & -0.01 \\
  \hline
 Very Low $\ell$ Pow. Spec. & 0.01 & 0.07 & -0.00 \\
  \hline
 High-$\ell$ Pow. Spec. & -0.23 & -0.58 & 0.15 \\
 \hline
\end{tabular}
\caption[]{\textit{The biases in the three cosmological parameters
    from the peak counts and power spectrum. The numbers in the table
    show the offset between the location of the best-fit in 3D and the
    correct fiducial cosmology along each of the 3 parameters
    $\sigma_8$, $w$, and $\Omega_m$.}}
\label{tab:Inferred bias}
\end{table}

\subsection{Bias directions and goodness-of-fit}
\label{subsec:UnderstandBias}

In this section, we address two important questions related to the
above results: (1) what is the quality of the biased best-fits? (2)
what determines the direction of the biases?  The first question is
especially important, since in principle, a poor best-fit can reveal
the presence of unaccounted-for parameters.

\begin{figure*}[htp]
\centering
\includegraphics[width=7 cm]{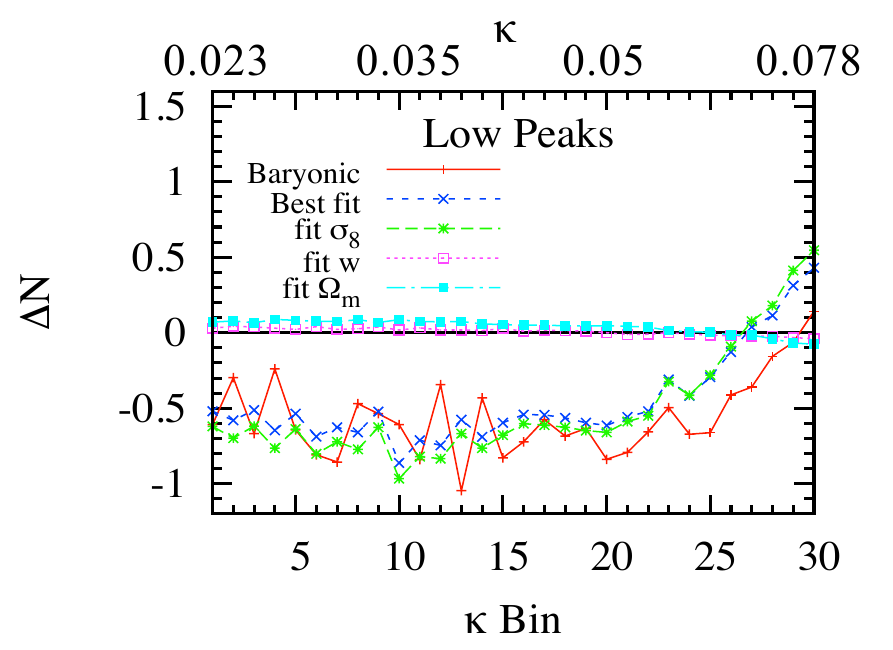}
\includegraphics[width=7 cm]{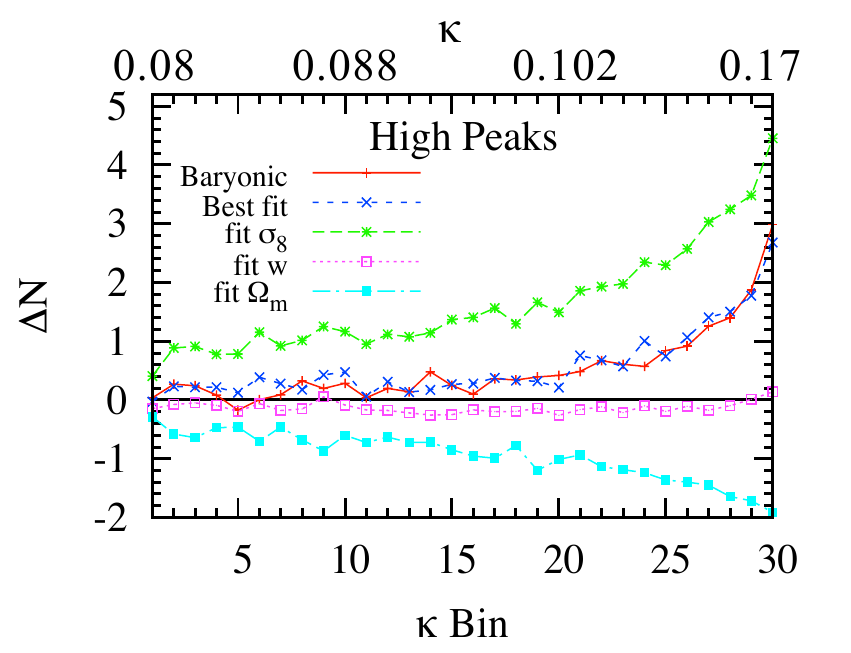} \\ \ \\
\includegraphics[width=7 cm]{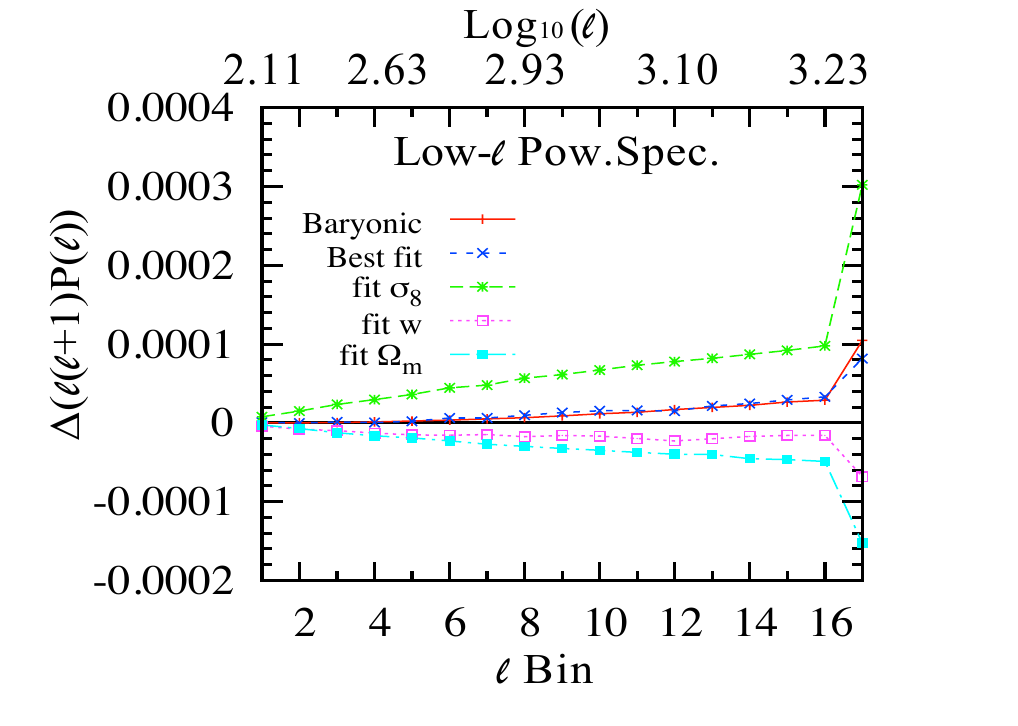}
\includegraphics[width=7 cm]{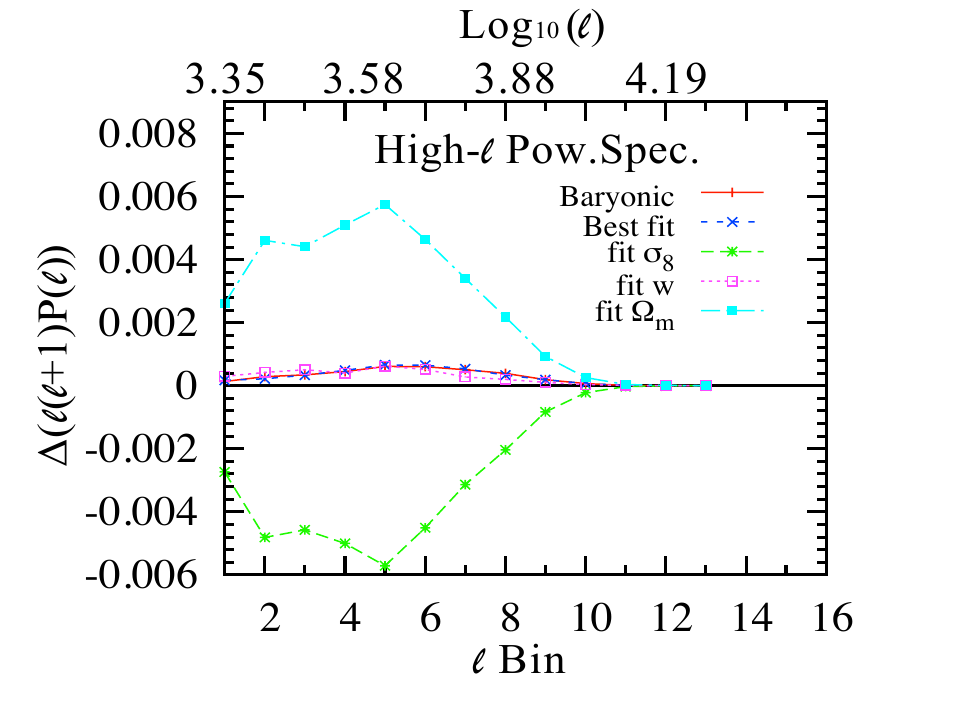}
\hfill
\caption[]{\textit{The figure compares the changes in the peak counts
    and the power spectra caused by the presence of baryons (solid red
    curves), and by the biased cosmological parameters in the best-fit
    baryonless models (dashed blue curves).  For every curve, the
    corresponding values in the fiducial, baryonless model is
    subtracted, so that only the offsets from this fiducial model are
    shown. The four panels (top left, top right, bottom left, bottom
    right) show the results for low peaks, high peaks, low--$\ell$
    power spectrum and high--$\ell$ power spectrum, respectively.  Bin
    numbers are shown on the bottom $x$ axis, with corresponding
    $\kappa$ and $\ell$ values indicated on the top of the figure
    (note that the binning is not linear in $\kappa$ and $\ell$; see
    \S~\ref{subsec:InferredBias} of the text for a full description of
    the binning scheme).  In general, the biased cosmologies are able
    to mimic the effect of baryons remarkably well.  The long-dashed
    [green], dotted [pink] and dot-dashed [cyan] curves show the
    ``decomposition'' of the biased best-fit cosmology among the three
    cosmological parameters: they show the effect of changing a single
    one of the three parameters $\sigma_8$, $w$, $\Omega_m$ at a time
    to its best fit value.  (The green, pink and cyan curves sum up to
    the blue curve.)  }}\label{fig:biasdecomp}
\end{figure*}

In Figure~\ref{fig:biasdecomp}, we show the peak counts and power
spectra in the baryonic model, compared directly to the best-fit
baryonless models.  The four panels (top left, top right, bottom left,
bottom right) show the results for low peaks, high peaks, low $\ell$
power spectrum and high $\ell$ power spectrum (following the
definitions of the low/high ranges in \S~\ref{subsec:InferredBias},
and using the corresponding best-fit points in Table~\ref{tab:Inferred
bias}).  In each case, the deviations from the baryonless fiducial
cosmological model are shown.  In each panel, the solid [red] curves
show the effect of the baryons, and the dashed [blue] curves show the
deviations that best mimic these curves, achieved in the best-fit
baryonless models.  As shown by the figure, in all four panels, we
find an excellent fit, i.e.\ the best-fit curves agree well with the
baryonic curves.  Although the best fit somewhat overpredicts the
number of peaks at the high-$\kappa$ end of the low peaks in the top
left panel, we find a total $\chi^2=30$. Since we have 30 bins, the
fit is good, with a reduced $\chi^2$ of unity (with somewhat better
fits in the other three panels).  This implies that when the baryonic
effects are neglected when fitting the data, the cosmological
parameters can be biased, and this will not be obvious from the low
quality of the fit.

To investigate the origin of the bias directions, the long-dashed
[green], dotted [pink] and dot-dashed [cyan] curves show the effect of
changing a single one of the three parameters $\sigma_8$, $w$,
$\Omega_m$ at a time, to its best fit value, i.e., the bias in one
parameter times the backward derivative of the observable with respect
to that.  (This means that the green, pink and cyan curves sum up to
the blue curve.)

In the case of the low peaks (top left panel), the best fit is driven
almost entirely by the bias in $\sigma_8$: the figure shows that the
change in $\sigma_8$, by itself, can mimic the baryonic effects quite
well.\footnote{Note that reducing the number of low peaks requires an
  {\em increase} in $\sigma_8$; this somewhat counterintuitive result
  has been found \cite{PCWLPC} and explained in detail ~\cite{Peaks}
  in our earlier work.}  Furthermore, the small residual would require
decreasing the number of peaks at the high-$\kappa$ end (in bins 20
and higher), while keeping the peak counts in the lower-$\kappa$ bins
unchanged.  Such a change in the shape of the peak count distribution
can not be accomplished by changing either $w$ or $\Omega_m$; as a
result, these two other parameters remain essentially unbiased.

The case of the high peaks (top right panel) is quite different, with
degeneracies between all three parameters playing an important role in
the bias.  In particular, the changes in the counts due to $\sigma_8$
and $\Omega_m$ have a strong degeneracy, so that these two parameters
work in the opposite direction.\footnote{This $\sigma_8-\Omega_m$
  degeneracy in the abundance of $\sim3\sigma$ peaks has been noted
  earlier \cite{DH10,Peaks}.}  However, the degeneracy is not perfect,
and leaves a nonzero residual -- interestingly, this residual can be
mimicked essentially perfectly by a relatively modest change in $w$.

Turning to the low-$\ell$ power spectrum (bottom left panel), the fit
is again driven by the bias in $\sigma_8$, but unlike in the case of
low peaks, $\sigma_8$ alone cannot mimic the baryonic effects. As a
result, $w$ and $\Omega_m$ play a role, working in the direction
opposite to $\sigma_8$, and producing a near-perfect fit.

The most interesting case is the high-$\ell$ power spectrum, shown in
the bottom right panel in Figure~\ref{fig:biasdecomp}.  Here the
degeneracy between $\sigma_8$ and $\Omega_m$ is so perfect that the
changes in these parameters (with opposite sign) essentially cancel
each other.  As a result, the net change in the high-$\ell$ power
spectrum can be attributed almost entirely to the change in $w$.  This
explains why there is a very large bias in $w$.  The fact that the bias
from the high peaks is driven by $\sigma_8$, whereas the bias from
the high-$\ell$ power spectrum ends up being dominated by $w$, makes it
less surprising that the biases from high peaks and the high $\ell$
power spectrum are in very different directions.

\section{Discussion}
\label{sec: Discussion}

\subsection{Robustness of the inferred bias}
\label{subsec: Robustness}

{\em How robust are the results to the number of bins?}  Our results
about biases have not been optimized over the number and placement of
bins. We have found, in general, that all the results listed in the
last section hold qualitatively, as long as the number of bins used is
not too small.  In Table~\ref{tab:number of bins}, we show the biases
of the three cosmological parameters $\sigma_8$, $w$, $\Omega_m$
inferred from the peak counts, with three different numbers of bins,
using noisy maps with $z_s=2$.  Bin boundaries were chosen such that
each bin contains the same number of peaks.  The table shows that the
biases from the low peaks are quite stable; as the number of bins is
varied from 20 to 30, the changes in the biases are always less than
the uncertainty on the bias ($\delta\sigma_8^\prime\approx 0.0065$,
$\delta w^\prime\approx 0.04$ and $\delta\Omega_m^\prime\approx
0.0065$ as mentioned above).  High peaks have larger changes in the
corresponding biases than the low peaks, but the sizes of these
fluctuations are still comparable to the uncertainties.  We conclude
that our qualitative results from the peak counts are robust as long
as $\gsim 20$ peak-height bins are used.

\begin{table}
\begin{tabular}{|c|c|ccc|} 
\hline
Observable  & bins & $\Delta\sigma_8$ & $\Delta w$ & $\Delta\Omega_m$\\
 \hline
 \multirow{3}{*}{Low Peaks} & 20 & 0.014 & 0.01 & -0.004 \\ 
  & 25 & 0.009 & 0.03 & 0.001 \\
  & 30 & 0.009 & 0.01 & -0.001 \\
  \hline
  \multirow{3}{*}{High Peaks} & 20 & 0.057 & 0.16 & -0.020 \\
  & 25 & 0.069 & 0.07 & -0.030 \\ 
  & 30 & 0.061 & 0.10 & -0.024 \\ 
 \hline
\end{tabular}
\caption[]{\textit{The inferred biases of the three cosmological
    parameters $\sigma_8$, $w$, $\Omega_m$ from the peak counts, using
    three different number of bins. Low and high peaks are always
    defined to be those with heights between $\sigma_{noise}\leq
    \kappa_{\rm peak}\leq 0.08$ and $\kappa_{\rm peak}\geq 0.08$; with
    $\sigma_{noise}=0.023$.  Bin boundaries were chosen such that each
    bin contains the same number of peaks.  }}
\label{tab:number of bins}
\end{table}

{\em How does the result depend on the finite difference derivatives?}
  To compute the inferred biases, the baryonic models were fitted with
  the non-baryonic models, using linear interpolation and backward
  finite difference derivatives.  Unless the model being evaluated is
  very close to one of the simulated cosmologies, there is no clear
  reason to prefer either the forward or the backward derivative. One
  concern is whether our results on the inferred biases change if we
  use different derivatives (or a 2nd order Taylor expansion).  To
  investigate this, we repeated our analysis using backward and
  ``2-sided" derivatives. In Table~\ref{tab: derivative}, we show the
  biases of the three cosmological parameters $\sigma_8$, $w$,
  $\Omega_m$ inferred from the peak counts, with all three types of
  derivatives.  As the table shows, the biases from the low peaks
  remain similar (or at least negligible, within errors) for each type
  of derivative.  However, high peaks have large and significant
  differences in the biases.

These difference may indicate a genuine asymmetry in the
cosmology-dependence of the peak counts, but the differences may be
partly a numerical artifact due to the $\sigma_8$--$\Omega_m$
degeneracy.  Given the cosmology-dependence of the peak counts (see,
e.g., Figure 8 in ref.~\cite{Peaks}), it is apparent that increases in
$\sigma_8$ can be mostly compensated by decreases in $\Omega_m$, and
vice versa (see also ref.~\cite{DH10}).  It is easy to imagine that as
a result of this degeneracy, small changes in the baryonic peak counts
that are being fit can produce large changes in the $\sigma_8$ and
$\Omega_m$ biases -- depending on where along the $\sigma_8-\Omega_m$
degeneracy curve the baryonic effect forces the best-fit (the
residuals could then also produce large changes in the $w$ bias).

In addition to the high peaks, we find that the biases from the
high-$\ell$ power spectrum also change by up to $30\%$ between using
backward vs.\ forward derivatives.  We conclude that, unfortunately, we
cannot determine the values of the biases of the high peaks and the
high-$\ell$ power spectrum accurately: we need a larger number of WL
maps to determine the derivatives with better precision, and the
parameter space needs to be sampled more finely with simulations to
map out the nonlinear dependence of the observables on the
cosmological parameters.  Nonetheless, our overall conclusion, namely
that the low peaks and the low-$\ell$ power spectrum are unbiased,
while high peaks and the high-$\ell$ power spectrum are noticeably
biased, holds for all types of derivative we have tried.  Furthermore,
it would be a remarkable coincidence if the true biases from the
latter two observables turned out to be in a degenerate direction; we
therefore also expect our generic conclusion about the possibility of
``self-calibration'' to remain valid.

\begin{table}
\begin{tabular}{|c|c|ccc|} 
\hline
Observable  & derivative & $\Delta\sigma_8$ & $\Delta w$ & $\Delta\Omega_m$\\
 \hline
 \multirow{3}{*}{Low Peaks} & backward & 0.009 & 0.01 & -0.001 \\ 
  & forward & 0.012 & 0.00 & -0.002 \\
  & 2-sides & 0.009 & 0.01 & --- \\
  \hline
  \multirow{3}{*}{High Peaks} & backward & 0.061 & 0.10 & -0.024 \\
  & forward & 0.035 & 0.05 & -0.009 \\ 
 & 2-sides & 0.026 & 0.09 & --- \\ 
 \hline
\end{tabular}
\caption[]{\it The biases of the three cosmological parameters
  $\sigma_8$, $w$, $\Omega_m$ inferred from the peak counts, using
  three different types of finite-difference derivatives.  The
  baryonic models were again fit by non-baryonic models, using linear
  interpolations with either backward, forward, or ``two-sided''
  derivatives, to make predictions for cosmologies in-between those
  that we simulated.  (The biases for $\Omega_m$ in the high peak case
  for the 2-sided derivative could not be determined reliable, owing
  to numerical issues related to the discontinuous switch in the
  derivative type at the fiducial $\Omega_m$.)}
\label{tab: derivative}
\end{table}

{\em How robust are the results to the number of realizations of the
baryonic maps?}  As mentioned above, our results are ultimately based
on the 200 baryonic maps in the fiducial cosmology.  To show that
these 200 baryonic maps are sufficient to capture the systematic
changes caused by the baryon cooling, in Table~\ref{tab:number of
realizations}, we compare the biases of the cosmological parameters
inferred from 100 vs.\ 200 realizations of baryonic maps.  We found
that the biases from 200 baryonic maps are close to the biases from
100 baryonic maps, with the differences of $<0.002$ in
$\Delta\sigma_8$, $<0.01$ in $\Delta w$ and $<0.002$ in $\Delta
\Omega_m$.  These differences are smaller than the uncertainties on
the biases discussed above. This is in reassuring contrast to the
apparently high demand on the accuracy of the derivatives, where not
even 1000 maps are enough to get stable biases for high peaks and the
high-$\ell$ power spectrum.  It is also worth noting that when we
tried to repeat our analysis using only 50 realizations, we found
significant differences in the biases. This shows that the baryonic
effects on the peak counts vary from one realization to another, and
at least $\gsim 100$ realizations are required to quantify them
reliably.  Interestingly, we have found that the baryonic effect on
the power spectrum is much more systematic, with little variation from
one map to another (see discussion below).

\begin{table}
\begin{tabular}{|c|c|ccc|} 
\hline Observable & \# of realizations & $\Delta\sigma_8$ & $\Delta w$
& $\Delta \Omega_m$\\ \hline \multirow{2}{*}{Low Peaks} & 100 &
0.011 & 0.010 & -0.001 \\
  & 200 & 0.009 & 0.005 & -0.001 \\
  \hline
  \multirow{2}{*}{High Peaks} & 100 & 0.059 & 0.101 & -0.023 \\
  & 200 & 0.061 & 0.103 & -0.024 \\ 
 \hline
\end{tabular}
\caption[]{\it The biases of the three cosmological parameters
$\sigma_8$, $w$, $\Omega_m$ inferred from the peak counts, based on
either 100 or 200 realizations of the baryonic maps.}
\label{tab:number of realizations}
\end{table}

{\em Effect of noise.}  An important question is how sensitive our
results are to the presence of galaxy shape noise. To address this
issue, we have repeated our analysis without adding noise to the maps.
Naively, one expects that noise might suppress the baryon-induced
differences in the peak counts. This expectation is confirmed by
Figure~\ref{fig:DiffHisto}, which shows, in particular, that the range
of peak heights that are unaffected is much narrower in the noiseless
case than in the noisy case.  We therefore proceed by re-defining the
low- and high-peaks in the noiseless maps to be those with heights
between $0\leq \kappa_{\rm peak}\leq \sigma_{0}$ and $\kappa_{\rm
peak}\geq \sigma_{0}$. Here $\sigma_{0}=0.022$ denotes the standard
deviation of the convergence $\kappa$ in the absence of noise, and
corresponds roughly to where baryon effects become significant (see
the red curve in Fig.~\ref{fig:DiffHisto}).

The biases of the cosmological parameters inferred from the low peaks
are found to be $(\Delta\sigma_8, \Delta w, \Delta\Omega_m)=(-0.005,
0.09, 0.001)$.  For high peaks the biases are $(\Delta\sigma_8, \Delta
w, \Delta\Omega_m)=(0.017, 0.24, 0.006)$.  These noiseless bias
vectors are generally different from those in the noisy case (rows
with 200 realizations in Table~\ref{tab:number of realizations}).  As
in the noisy case, the biases from the low peaks are still much
smaller than from the high peaks (we have verified that this is also
true for the forward and 2-sides derivatives).  However, the simple
expectation that noise reduces the bias holds only for both low- and
high peaks for $w$.  The noise {\em increases} the bias in $\sigma_8$
for both types of peaks.  Finally, the $\Omega_m$-bias is decreased by
noise for low peaks, but increased for high peaks.  In our analysis,
we have assumed that the noise is Gaussian, and is known perfectly.
The above changes in the bias vectors imply that precise measurements
of the shape noise will indeed be crucial when attempting to model
baryon effects in the WL data.

We have also found that the biases in the noiseless maps are not as
stable with respect to the number of realizations as in the noisy
case.  The worst case from backward derivative is the noiseless low
peaks.  For example, when we use 100 realizations, the biases change
by a factor of $\approx$two, to $(\Delta\sigma_8, \Delta w,
\Delta\Omega_m)=(-0.011, 0.05, 0.003)$.  As we will see in the next
paragraph, this is in contrast with the behaviour of the bias from the
power spectrum, which is always very stable to number of realizations.
A possible explanation of this difference is that changing the
concentration parameter can impact any individual peak either
way. This is because the change in the contribution $\kappa$ to a peak
from an individual halo can be positive or negative, depending on the
impact parameter of the halo; halos contributing (especially to the
low) peaks have a broad range of impact parameters (these points will
be demonstrated in Figure~\ref{fig:Diffconv} below).  This effect is
reduced in the presence of noise, since noise itself tends to give the
largest contribution of $\kappa$ to low peaks.

{\em Robustness of biases from the power spectrum.}  As mentioned
above, we have found that the results from the power spectrum, both
with and without noise, are always very robust with respect to the
number of realizations of baryonic maps.  The small sizes of the error
bars in Figure~\ref{fig:DiffPs} already show that baryonic effects
have much weaker map-to-map variations, compared to those in
Figure~\ref{fig:DiffHisto}.  The differences in the biases inferred
from 200 vs.\ 100 realizations are less than 0.001 in
$\Delta\sigma_8$, 0.01 in $\Delta w$ and 0.001 in $\Delta\Omega_m$.
In fact, we have found that for the power spectrum at $\ell\geq2000$,
even a {\em single} map is sufficient to determine the biases with
$\lsim 10\%$ accuracy.  For the power spectrum at $\ell\leq2000$,
which is numerically somewhat less stable than the high-$\ell$ power
spectrum, we have found that $\gsim 5$ maps are needed to achieve the
same accuracy.  These numbers refer to the noisy maps; in the
noiseless case, the $\ell\leq2000$ power spectrum would require $\gsim
10$ maps.

The biases from the power spectra are more sensitive to derivative
types, with absolute changes comparable to those for the peaks.
However, compared to peaks, the biases from the power spectra are
larger to begin with, making the biases more stable in a fractional
sense.  The differences in biases from the power spectra with and
without noise are $<30\%$; the only exception is the noisy case and
the high-$\ell$ range, for which there is a change by a factor of 2
in $\Delta w$.  We have studied the sensitivity to the number of bins
only for the high-$\ell$ case (this is because we did not have a
sufficient number of pre-bins saved in the low and very low-$\ell$
ranges to increase the resolution in those cases).  In our fiducial
set of computations above, we used 13 high $\ell$ bins.  We found that
as we increase the number of bins from 13 to 15 to 20, the changes in
the biases are comparable to those for the high peaks, with the
13-to-15-change being a bit smaller than the 13-to-20-change.
However, again because the power spectrum in general has much larger
absolute biases, even in the worst case, changing from 13 to 20 bins,
the corresponding fractional change in the parameter biases is low
($\lsim15\%$).

\subsection{Cosmology from low-bias statistics}
\label{subsec:Constraint}

We have shown that baryon cooling causes a strong increase in the
number of high peaks and the small scale power spectrum.  The number
of low-amplitude peaks and the large-scale power spectrum are both
relatively insensitive to baryon cooling, and thus deliver nearly
bias-free cosmological constraints. Figure~\ref{fig:Inferred bias}
reveals that the area of the marginalized 2D error contours from these
low-biased statistics are roughly comparable to those from their more
biased counterparts.  In this section, we will further explore how much
of the cosmological sensitivity resides in the low-bias statistics.

To assess the fraction of the cosmological sensitivity coming from low
peaks and large-scale power spectra, we first computed a
$\Delta\chi^2$, analogous to the quantity defined in Eq.~(\ref{chi2})
above, except using the difference $\Delta N$ between the mean number
of peak in pairs of CDM simulations with different $\sigma_8$, $w$ and
$\Omega_m$.  This measures the significance of the differences caused
by the changes in the individual cosmological parameters.  We find
that the $\Delta\chi^2$ values from the relatively unbiased
low-amplitude peaks exceed half of the $\Delta\chi^2$ values obtained
from all peaks.  On the other hand, the $\Delta\chi^2$ values from the
large-scale power spectrum are typically reduced by a factor of
$\sim5$ compared to using the power spectrum on all scales (the
exception is $w$, for which the large angular scales contain $\approx$
half of the total $\Delta\chi^2$).

The above shows that the low peaks contain most of the raw
cosmological sensitivity to each individual parameter, but it neglects
degeneracies among parameters, which ultimately drive the marginalized
constraints.  We therefore next assess the fraction of the full
cosmological sensitivity, coming from low peaks and large-scale power
spectra, when all three parameters are simultaneously varied. For this
purpose, we repeated our analysis as outlined in this paper, except we
bootstrapped the CDM maps from our 45-simulation map set, rather than
the baryonic maps.  This yields constraint in the three-dimensional
parameter space $\sigma_8$, $w$ and $\Omega_m$. We measure the
three-dimensional volumes in the parameter space containing 68\% of
the probability for all peaks and low peaks separately, and likewise
for the very low-$\ell$ and the full power spectra. The figure of
merit (FOM) is defined to be the inverse of these volumes, and
indicates the overall strength of the constraints. The ratio of the
FOM from low vs. all peaks and the very low-$\ell$ vs. the full power
spectra then captures the ``fraction'' of the full cosmological
sensitivity in the unbiased statistics. This ratio can be converted
roughly into a per-parameter estimate by taking the third root,
although the differences in constraints on individual parameters would
of course depend on the exact shape of the constraint volume. We list
these quantities in Table~\ref{tab:chisq low and total}.

\begin{table}
\begin{tabular}{|c|c|c|c|c|} 
\hline
Observable & FOM    & FOM  & Fraction & Fraction \\
           & (all)  & (low-bias) & (low/all)  & per parameter \\
\hline
Peaks & $1.30\cdot10^6$ & $0.723\cdot10^6$ & $0.56$ & $0.82$ \\ 
\hline
Pow. Spec. & $1.10\cdot10^6$ & $0.060\cdot10^6$ & $0.055$ & $0.38$ \\
\hline
\end{tabular}
\caption[]{\textit{The strength of the cosmological constraints for
peak counts and power spectrum, expressed in terms of a figure of
merit (FOM; defined as the inverse of the three-dimensional error
volume in the $\sigma_8, \Omega_m, w$ parameter space).  FOM values
are compared for the relatively unbiased low-amplitude peaks and very
low-$\ell$ ($100<\ell<1000$) power spectrum (second column) vs.\ all
peaks and the power spectrum on all scales (first column).  The third
column indicates the ratios of these FOM (indicating the fraction of
the cosmological constraints contained in the low-bias ranges of both
statistics), while the rightmost column shows the cube root of this
ratio (a rough estimate of the fraction of the constraint on
individual parameters).}}
\label{tab:chisq low and total}
\end{table}

As is evident from the table, the peaks and the power spectrum deliver
comparable FOMs when the whole range is used. However, the essentially
unbiased low-amplitude peaks contain more than half (a fraction
$0.56$) of the FOM, whereas this fraction is only $\sim 0.05$ for the
very low-$\ell$ power spectrum (the latter increasing to $\sim 0.16$
if multipoles up to $\ell<2000$ are included). In terms of the
fraction per parameter, the low peaks contain approximately 82\% of
the information, whereas the very low-$\ell$ power spectrum contains
38\% (the latter number increasing to 54\% with multipoles up to
$\ell<2000$).  We conclude that the low-bias statistics contain the
(slight) majority of the cosmological sensitivity for peaks, but a
relatively small fraction for the power spectrum.

\subsection{The insensitivity of low peaks to baryon cooling}
\label{subsec:MProbust}

We next examine why the number of low peaks is insensitive to baryon
cooling, as modeled by an increase in the halo concentration
parameters.  In short, we attribute the lack of sensitivity to two
reasons: halos contributing to low peaks have large off-sets from the
line-of-sight toward each peak (and therefore the light rays do not
``see'' the halo cores) and also relatively low masses (so that the
details of the projected halo density profiles are ``washed out'' by
the 1-arcmin smoothing).  We demonstrate these points explicitly
below.

\begin{figure}[htp]
\centering
\includegraphics[width=8 cm]{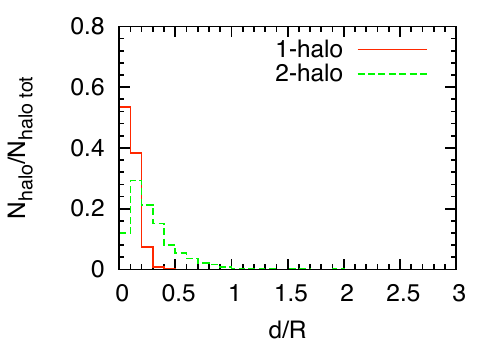}\\ 
\includegraphics[width=8 cm]{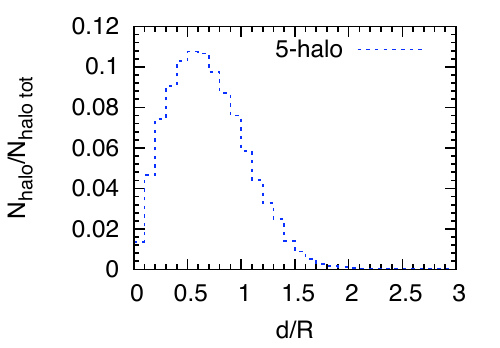}\\ 
\hfill
\caption[]{\textit{The distribution of the impact parameters (in units
of the virial radius) at which light rays, corresponding to the
centers of WL peaks, intersect dark matter halos along the line of
sight.  The upper panel is for high peaks, which are typically
produced by one or two halos.  The panel shows, separately, the
impact-parameter distribution for the dominant halos for 1-halo peaks
(solid red curve), and for 2-halo peaks (dashed green curve).
``$n$-halo peaks'' are defined to be those for which the sum of $n$
halos along the line of sight accounts for at least $50\%$ of the halo
contribution to the total peak convergence.  The light rays typically
go near the halo centers, with impact parameters $<0.2R_{180}$.  In
contrast, for the low peaks, shown in the lower panel, the light rays
have a large impact parameter.  These peaks are typically produced by
4-8 halos; the panel shows the distribution for the typical case of
5-halo peaks, which has a maximum at $\approx 0.6R_{180}$.}}
\label{fig:impactd}
\end{figure}

In Figure~\ref{fig:impactd}, we show the distribution of the impact
parameters (in units of $R_{180}$) at which light rays, corresponding
to the centers of WL peaks, intersect dark matter halos along the line
of sight. Following ref.~\cite{Peaks}, for each peak, we identify all
halos along the sightline, and rank them by their contribution to the
peak's height. Going down this ranked list (beginning with halos with
the largest contribution), we sum the $\kappa$ contributions from the
first $n$ halos until these $n$ halos account for $\geq 50\%$ of the
total halo contribution to the peak convergence.  This procedure
assigns the number $n$ to each peak (hereafter referred to as an
``$n$-halo peak'').

The top panel relates to high peaks (with heights of $\kappa \geq
0.08$), which are typically produced by one or two halos. The panel
shows, separately, the impact-parameter distribution for the dominant
halos for 1-halo peaks, and for 2-halo peaks.  Low peaks (with heights
between $\sigma_{\rm noise}\leq \kappa_{\rm peak}\leq 0.08$; with
$\sigma_{noise}=0.023$) are typically produced by 4-8 halos. The
representative case of 5-halo peaks is shown in the bottom panel.  Our
baryonic model, with a 50\% boost in the concentration parameter, was
used for this figure (with source galaxies at redshift $z_s=2$ and
with galaxy shape noise and 1 arcmin smoothing included as before).

As the figure shows, for high peaks, the light rays typically go near
the halo centers, with impact parameters $< 0.2R_{180}$. In contrast,
for low peaks, the light rays have large impact parameters, with a
maximum near $\approx 0.6R_{180}$.  This is as expected: as more halos
contribute to a peak, the contribution from each halo is lower, and
the halos are less precisely aligned along the line of sight.

\begin{figure}[htp]
\centering
\includegraphics[width=8 cm]{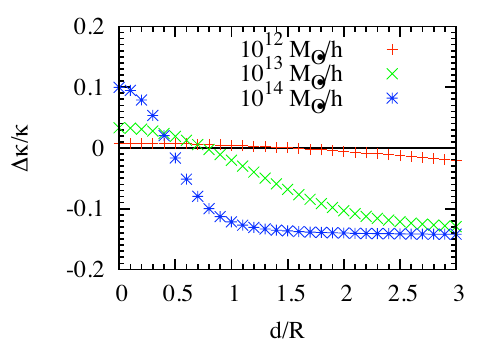}\\ 
\hfill
\caption[]{\textit{The figure shows the fractional difference in the
convergence contribution from halos, caused by a 50\% increase in
their concentration, as a function of the impact parameter $d$ (in
units of the halo's virial radius $R_{180}$).  The red plus signs,
green crosses, and blue snowflakes correspond to halos with masses of
$10^{12}$, $10^{13}$, and $10^{14}h^{-1} M_\odot$, as labeled.  The
halos and source galaxies are located at redshift $z=0.5$ and $z_s=2$,
respectively and we assume the halos have NFW profiles, with a
concentration parameter (before the 50\% boost) given by
Eq.~(\ref{eq:NFW4}).  Lower-mass halos are less affected by the boost
in concentration.  These halos are more compact, and the 1 arcmin
smoothing makes their convergence contribution less sensitive to their
intrinsic density profile.}}
\label{fig:Diffconv}
\end{figure}

Figure~\ref{fig:Diffconv} shows the fractional difference in the
convergence contribution from halos, caused by a $50\%$ increase in
their concentration, as a function of the impact parameter (in units
of the halo's virial radius $R_{180}$).  The three curves correspond
to halos with masses of $10^{12}$, $10^{13}$, and $10^{14}h^{-1}
M_\odot$, as labeled.  The halos and source galaxies are located at
redshift $z=0.5$ and $z_s=2$, respectively and we assume the halos
have NFW profiles, with a concentration parameter (before the 50\%
boost) given by Eq.~(\ref{eq:NFW4}). As the figure shows, lower-mass
halos are less affected by the boost in concentration.  These halos
are more compact, and the 1 arcmin smoothing makes their convergence
contribution less sensitive to their intrinsic density profile.

Figures~\ref{fig:impactd} and~\ref{fig:Diffconv} together allows us to
assess the sensitivity of peak counts to baryons for both high and low
peaks. High peaks are dominated by halos with higher masses $10^{13} -
10^{14}h^{-1} M_\odot$, and the dominant halos for the 1-halo peaks
have masses near the upper end of this mass range. On the other hand,
the low peaks are dominated by halos with lower masses, between
$10^{12} - 10^{13}h^{-1} M_\odot$. Combining the results from
Figures~\ref{fig:impactd} and \ref{fig:Diffconv}, we see that high
peaks with masses $\lesssim 10^{14}h^{-1} M_\odot$ and impact
parameters $< 0.2R_{180}$ have typical fractional differences of about
8 percent.  By comparison, low peaks with masses $10^{12} -
10^{13}h^{-1} M_\odot$ and impact parameters $\approx0.6R_{180}$, have
typical fractional differences of about 2 percent, much lower than
high peaks. This explains why the low peaks are less sensitive to
baryon cooling.

\section{Summary and Conclusions}
\label{sec:conclusions}

In this paper, we have studied the impact of the cooling and
concentration of baryons in dark matter halos on WL observables, by
manually steepening the density profile of each DM halo in a suite of
ray-tracing N-body simulations. Our main findings are that a) the low
peaks ($0.023 \leq \kappa_{peak} \leq 0.08$) are mostly unaffected, b)
high peaks ($\kappa_{peak} \geq 0.08$) and the power spectrum at
$\ell>1,000$ lead to biases if baryonic effects are neglected in the
cosmological parameter estimation, c) the bias in high peaks is
comparable to the bias for the low-$\ell$ ($\ell<2000$) power
spectrum, and d) the high-$\ell$ ($2000<\ell<20,000$) power spectrum
exhibits a much stronger bias in a different direction.

We find a large increase in the amplitude of the small-scale power
spectrum, caused by the steepening of halo profiles, confirming
previous works.  It is unsurprising that we find a corresponding large
increase in the number of high peaks, since these peaks are typically
caused by individual halos.  However, the fact that low peaks are
essentially insensitive to an increase in the halo concentration was
surprising (at least to us).  This is an especially important finding,
since these low peaks contain the majority of the cosmological
information contained in the entire set of peaks.  We attribute the
robustness of these low peak to the fact that they are created by an
alignment of typically 4--8 of low-mass ($\sim 10^{12}-10^{13} {\rm
  M_\odot}$) halos \cite{Peaks}, with each halo typically offset from
the line-of-sight towards the peak by a fair fraction of the virial
radius ($\sim 0.5 R_{\rm vir}$). As a result, light-rays corresponding
to the peak do not pierce the cores of these halos, where baryonic
effects are most significant.  Additionally, the halos contributing to
the low peaks are compact on the sky. As a result, when their
intrinsic profiles are convolved with the smoothing filter (with an
angular scale of 1 arcmin in our case), applied to the raw convergence
maps, the details of the profiles are washed out.

We have explicitly computed the biases in the cosmological parameters
$w$, $\sigma_8$ and $\Omega_m$ when peak counts and power spectra are
fit neglecting the baryonic effects. We find that the biases from the
low peaks and from the very low-$\ell$ power spectrum ($\ell<1000$)
are small (consistent with zero within errors). The biases from high
peaks and from the low-$\ell$ power spectrum ($\ell<2000$) are
significant and comparable both in magnitude and sign (e.g. both
biases are $\Delta w\approx +0.1$ for dark energy). It is an important
result that the high peaks are no more biased than the low-$\ell$
power spectrum ($\ell<2000$), the latter being the range of scales in
the baseline plans by the LSST WL survey, to infer cosmological
parameters~\cite{LSST:2009pq}.

Our finding that biases from the high-$\ell$ power spectrum
($2000<\ell<2\times10^4$) are very large (between $\Delta w=-0.3$ and
$-0.6$), but in a very different direction in the ($w$, $\sigma_8$,
$\Omega_m$) parameter space than the bias from high peaks, also has an
important general implication. It suggests the possibility of an
effective self-calibration, by tuning the concentration parameter to
bring the different contours in the cosmological parameter space into
alignment. However, given the large magnitude of this bias, and
keeping in mind that additional unknown ``baryonic parameters'' will
be present in a fully realistic case, it is difficult to see how this
self-calibration could be achieved to the sub-percent accuracy level
targeted by LSST.  We note, however, that biases from the power
spectrum extended down to $\ell<2,000$ are lower, and still in a
modestly different direction than that of the high peaks, possibly
allowing more accurate self-calibration.  Until baryonic effects are
much better understood, it may be advantageous to restrict the
analysis to statistics that are nearly unbiased, such as the low peaks
and the very low ($\ell<1,000$) power spectrum we identify here.  Our
result suggests than in the case of peaks, less than half of the
cosmological constraining power may be lost.

Our two main qualitative results are likely robust: constraints from
sufficiently low-amplitude peaks will be relatively unbiased; and some
form of self-calibration will likely be realized in a real survey.
However, our study is clearly highly idealized. We have identified a
number of caveats arising from our toy model for the impact of
astrophysical processes, our limited number of simulations (both in
terms of sampling the cosmological parameter space and in terms of the
number of realizations), our simplified treatment of source galaxies,
noise, and our neglect of all instrumental errors.  Our results call
for follow-up work with more extensive simulations and more realistic
modeling, to address these shortcomings.

\begin{acknowledgments}

  This research utilized resources at the New York Center for
  Computational Sciences, a cooperative effort between Brookhaven
  National Laboratory and Stony Brook University, supported in part by
  the State of New York. This work is supported in part by the
  U.S. Department of Energy under Contract No. DE-AC02-98CH10886 and
  by the NSF under grant AST-1210877.  JMK and KMH acknowledge support
  from NASA's Jet Propulsion Laboratory through Contract 1363745. The
  simulations and non-baryonic WL maps were created on the IBM Blue
  Gene/L and /P New York Blue.  The baryonic maps were created and the
  analyses were performed on the LSST/Astro Linux cluster at BNL.

\end{acknowledgments}

\bibliographystyle{physrev_mod} 

\end{document}